\documentclass{sig-alternate-05-2015}
\usepackage{graphicx}
\usepackage{amssymb, amsmath}
\usepackage{epstopdf}
\usepackage{url}
\usepackage{caption}
\usepackage{subcaption}
\usepackage{multirow}
\usepackage{lipsum}
\usepackage{tabularx}

\usepackage{enumitem}
\usepackage{makecell}
\usepackage{marginnote}
\usepackage{flushend} 
\usepackage{graphics} 
\usepackage{times}
\usepackage{epstopdf}
\usepackage{array}
\usepackage{booktabs} 
\usepackage{amssymb,latexsym,amsmath}    
\usepackage{booktabs}
\usepackage{multirow}
\usepackage{booktabs, multicol, multirow}
\usepackage{rotating}
\usepackage{tikz}
\usepackage{cite}
\usepackage{subfig}
\usepackage{color, colortbl}
\definecolor{LRed}{rgb}{1,.8,.8}
\definecolor{LGreen}{rgb}{0.8,1,0.8}
\definecolor{HRed}{rgb}{1,.2,.2}
\definecolor{Yellow}{cmyk}{0,0,0.5,0}
\definecolor{Gray}{gray}{0.85}
\usepackage{array}
\usepackage{booktabs} 
\usepackage{amssymb,latexsym,amsmath}     
\usepackage{booktabs}
\usepackage{multirow}
\usepackage{booktabs, multicol, multirow}
\usepackage{rotating}
\usepackage{graphicx}
\usepackage{epstopdf}
\usepackage{tikz}
\usepackage{url}
\usepackage{bigstrut}
\usepackage{colortbl}
\usepackage{hhline}
\usepackage{bigstrut}
\usepackage{tabu}
\newcommand{\subparagraph}{}
\usepackage{gensymb}
\usepackage{titlesec}
\usepackage{balance}

\def \sys {\textit{TransitLabel}}

\begin{document}

\title{TransitLabel: A Crowd-Sensing System for Automatic  Labeling of Transit Stations Semantics}

\numberofauthors{1} 
\author{
Moustafa Elhamshary$^{\dag\ddag}$,Moustafa Youssef$^{\dag}$,Akira Uchiyama$^{\ddag}$,Hirozumi Yamaguchi$^{\ddag}$,Teruo Higashino$^{\ddag}$\\
\affaddr{$\dag$ Wireless Research Center, Egypt-Japan University of Science and Technology (E-JUST), Alexandria, Egypt.}\\
\affaddr{$\ddag$  Graduate  School of Information  Science and Technology, Osaka University, Suita, Osaka, Japan.}\\
\affaddr{\{mostafaelhamshary, moustafa.youssef\}@ejust.edu.eg, \{uchiyama, h-yamagu, higashino\}@ist.osaka-u.ac.jp}
}

\CopyrightYear{2016} 
\setcopyright{acmcopyright}
\conferenceinfo{MobiSys'16,}{June 25-30, 2016, Singapore, Singapore}
\isbn{978-1-4503-4269-8/16/06}\acmPrice{\$15.00}
\doi{http://dx.doi.org/10.1145/2906388.2906395}

\maketitle
\begin{abstract}

We present \sys{}, a  crowd-sensing system for automatic enrichment of  transit stations indoor floorplans  with different semantics like  ticket vending machines, entrance gates, drink vending machines, platforms, cars' waiting lines, restrooms,  lockers, waiting (sitting) areas, among others. Our  key observations show  that  certain passengers' activities (e.g., purchasing  tickets, crossing entrance gates, etc)  present   identifiable signatures on one or more cell-phone sensors. \sys{}  leverages this fact to automatically and unobtrusively  recognize different passengers' activities, which in turn are mined to infer their uniquely  associated stations semantics. Furthermore, the locations of  the discovered semantics are automatically estimated from the inaccurate  passengers' positions  when these semantics are identified.\\
We  evaluate \sys{} through a field experiment in eight different train stations  in Japan. Our results show that \sys{} can detect  the fine-grained stations semantics accurately with  7.7\% false positive rate and 7.5\% false negative rate on average. In addition, it can  consistently  detect the location of discovered semantics accurately, achieving an error  within 2.5m on average for all semantics. Finally, we show that \sys{} has a small energy footprint on cell-phones, could be generalized to other stations, and is robust to different phone placements; highlighting its promise as a ubiquitous  indoor maps enriching service.

\end{abstract}

\section{Introduction}

With the fact that people spend most of their time at indoor spaces,  indoor  Location Based Services (LBSs)  are being developed at a phenomenal   rate with a variety of applications  including mapping and navigation services, point-of-interest finders, geo-social networks, and advertisements. With the fact that people spend most of their time
at indoor spaces, indoor LBSs are being developed at a phenomenal
rate. A key requirement to  indoor LBSs is the availability of indoor maps  to display the user location on. These LBSs have still a huge
potential for enhancement if rich semantic information is attached to indoor maps to support
a wide class of indoor mapping applications (especially large
public buildings that are visited daily by many people like railway stations,
airports, museums, etc). Realizing the economic value of this technology,  a number of commercial navigation systems for indoor mapping have started to emerge. In late 2011, Google Maps started to expand its coverage by providing detailed floorplans   for a few malls and airports in the U.S. and Japan as well as allowing buildings owners around the world to upload their indoor floorplans. Nevertheless,  these maps are still  limited in coverage  to a small number of countries featuring only some major airports, shopping malls,  etc. This  limitation in coverage  is due  in part to  the following reasons: (1) buildings owners  may not allow sharing of their floorplans in public for privacy reason, (2) buildings internal structures often evolve over time, and/or (3) manual creation of these maps requires slow,  labor-intensive tasks, and they are subject to intentional incorrect data entry by malicious users.\\
Railway stations, as an example of indoor places, are  a key part of the day-to-day lives of people  having  millions of passengers every day (e.g., Shinjuku station in Japan  has  3.64 million passengers/day on average\footnote{https://en.wikipedia.org/wiki/Shinjuku\_Station\label{shinj}}). In highly populated countries, major stations have large indoor spaces (e.g., Shinjuku  station in Japan has 36 platforms and over 200 exits\textsuperscript{\ref{shinj}}). Therefore, a number of indoor navigation apps, e.g. the Tokyo station underground area navigation app\footnote{http://en.rocketnews24.com/2016/02/17/tokyos-busiest-train-stations-have-a-new-free-english-compatible-navigation-app/}, for stations have started to emerge. These applications, however, are built upon a manually created map of the building showing all important points of interest (e.g., fare collection gates, ticket vending machine, etc), which impedes their scalability to large scale deployments at different stations.
 For example, Google indoor maps  covers less than 50 transit stations worldwide, which are a small fraction of the thousands of stations on Earth\footnote{https://support.google.com/gmm/answer/1685827?hl=en}. The lack of detailed digital floorplans for railway stations highlighting  locations of various semantics limits passengers' experience, especially for foreigners or first-time visitors. Consequently, this sparks the need for the automatic construction of detailed indoor floorplans for transit stations. \\
 \begin{figure*}[!t]
        \centering
        \begin{subfigure}[b]{0.23\textwidth}
                \includegraphics[width=1\textwidth,height=3cm]{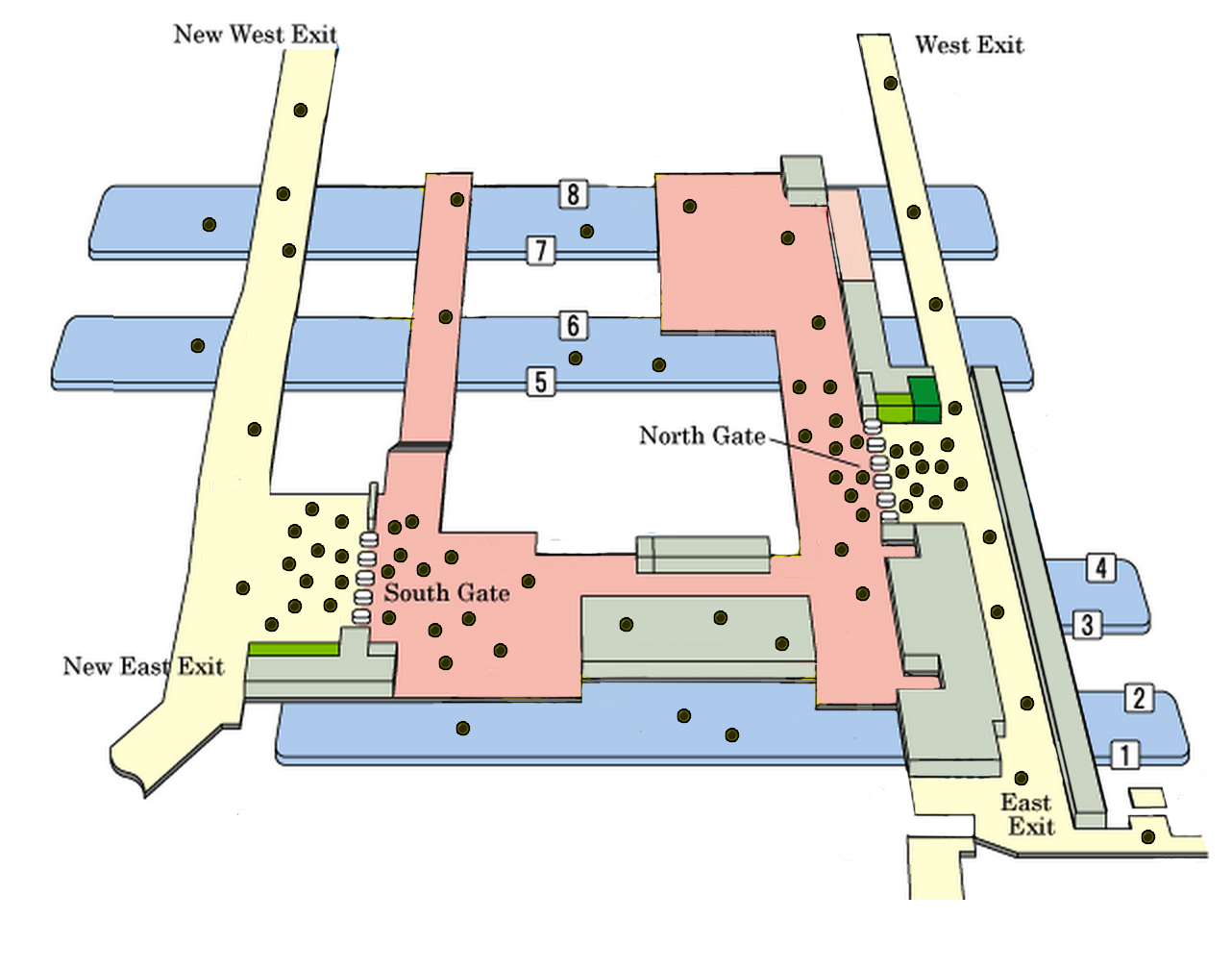}
           \caption{The locations of passing through fare collection gate  activity  as estimated by pedestrian dead-reckoning (PDR) are highlighted on  the floorplan.}
       %    \label{transport}s
        \end{subfigure}%
        \quad
        \begin{subfigure}[b]{0.23\textwidth}
                \includegraphics[width=\textwidth,height=3cm]{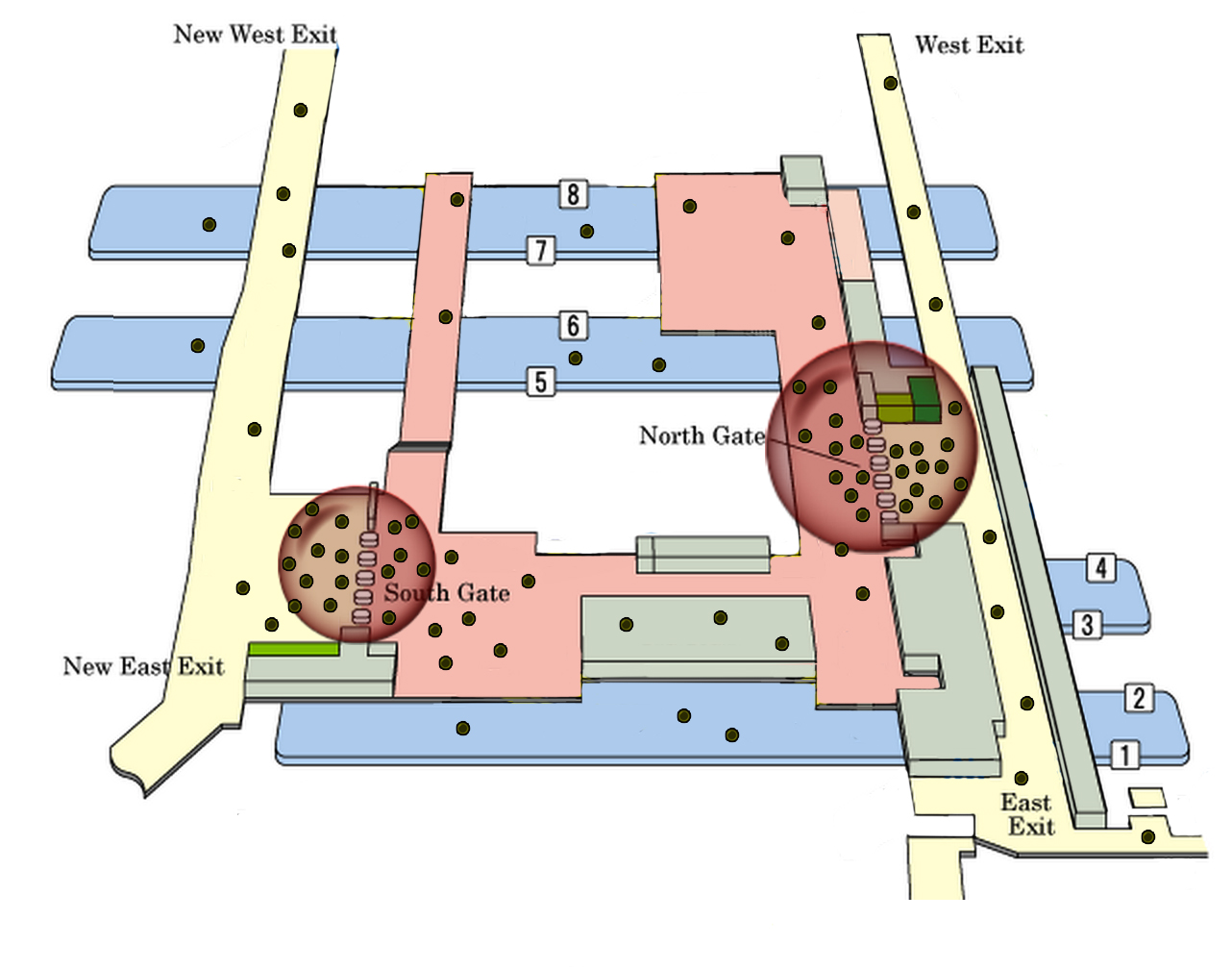}
              \caption{The output clusters  from the DBSCAN  based on crossing fare collection gates activity locations are highlighted on the floorplan.}
          %    \label{other}
        \end{subfigure}
        \quad
         \begin{subfigure}[b]{0.23\textwidth}
                \includegraphics[width=1\textwidth,height=3cm]{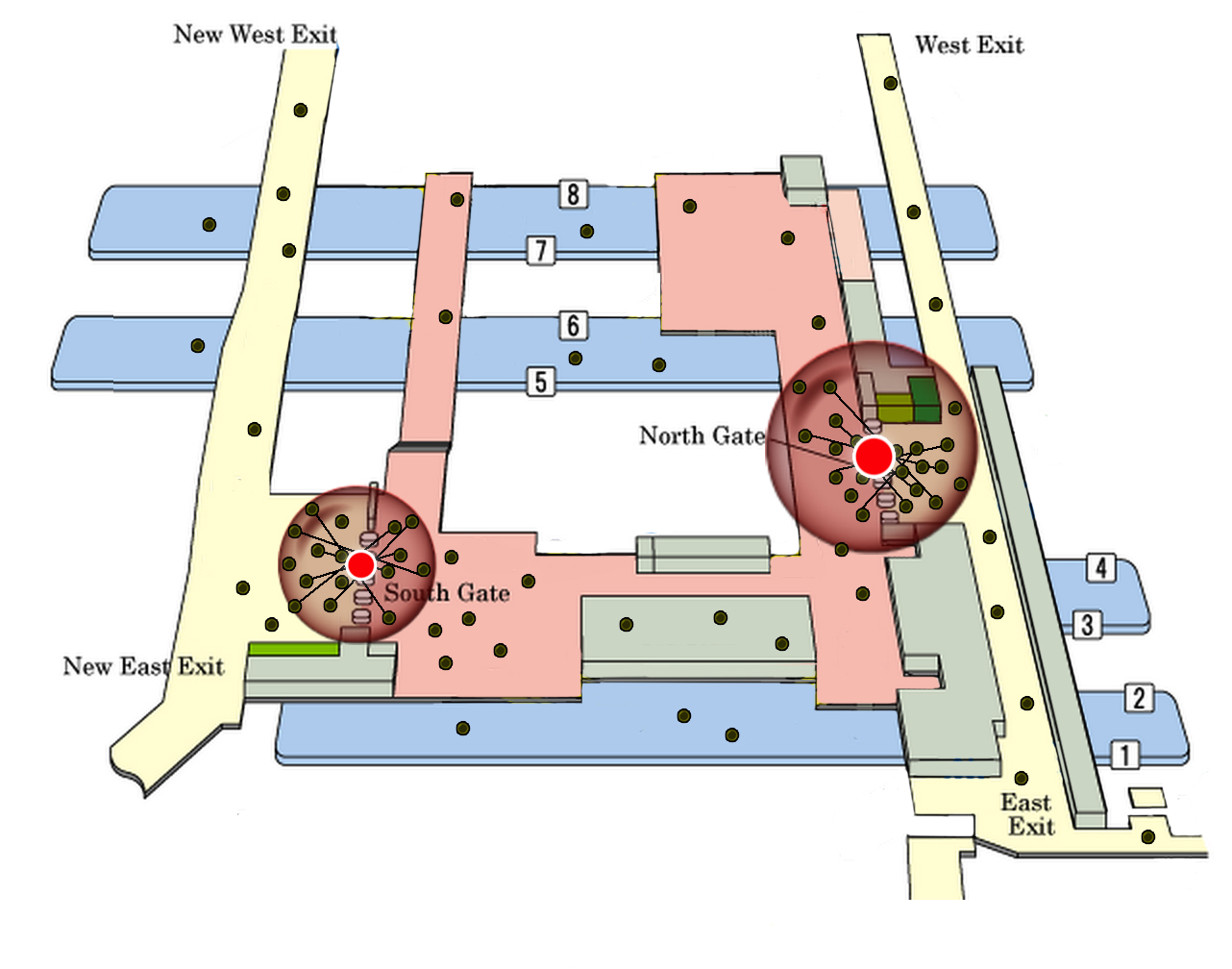}
           \caption{The semantics locations (e.g., fare collection  gates)  are estimated as the center of mass of all samples within the output clusters.}
      %     \label{transport2}
        \end{subfigure}%
        \quad
        \begin{subfigure}[b]{0.23\textwidth}
                \includegraphics[width=\textwidth,height=3cm]{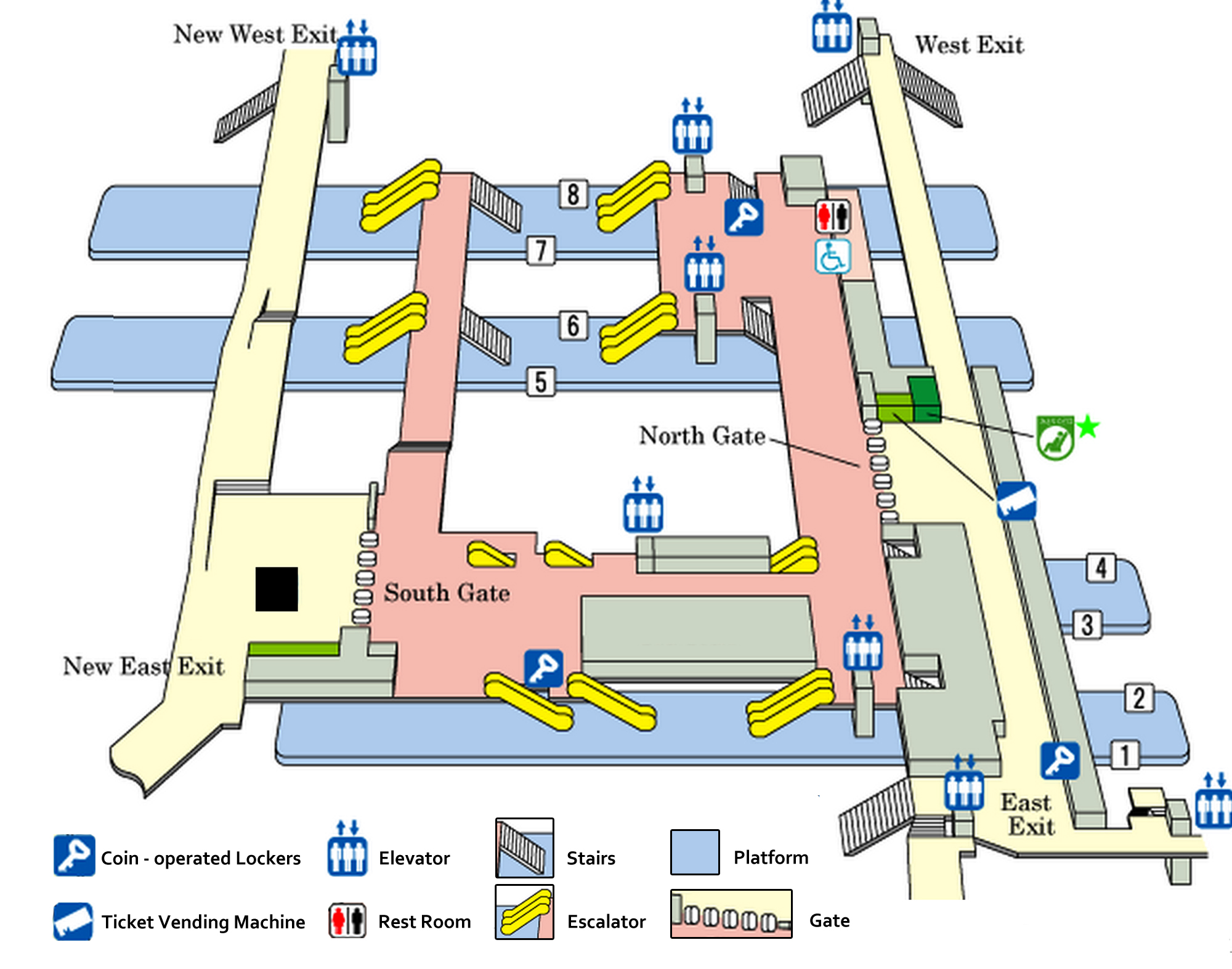}
              \caption{The station indoor map with the discovered semantics  locations, estimated by \sys{}, are highlighted on the floorplan (\sys{} output).}
            %  \label{other2}
        \end{subfigure}
        \caption{An example of  \sys{} in action  to identify the location of ticket  gates in a railway station. }
             \label{eg}
\end{figure*}
To resolve  this problem, the research community recently has embarked to address the problem of automatic construction of indoor floorplans by  exploiting motion trajectories of mobile phone users  \cite{alzantot2012crowdinside,jiang2013hallway,gao2014jigsaw}. These systems proved the feasibility of estimating the general layout of a building  \cite{alzantot2012crowdinside,jiang2013hallway,gao2014jigsaw}, identifying rooms shape and dimensions  \cite{alzantot2012crowdinside,gao2014jigsaw}, along with identifying other points of interest such as store entrances \cite{alzantot2012crowdinside,gao2014jigsaw}. Nevertheless,  none of these approaches provide semantic-rich floorplans  where various semantics are tagged on the floorplan  that are necessary for many of today's  map-based applications. For example,  stations indoor navigation systems should  rely on important semantics to better guide passengers to their destinations; a station evacuation planning is ineffective if maps are not tagged with  emergency exit stairs;  a person with disability needs a map showing  elevator-enabled routes; and an occasional passenger needs a map of important semantics that she must use to board the train (e.g., ticket gates, etc). Moreover, the discovered semantics can be leveraged to provide accurate calibration-free indoor localization, by providing opportunities for dead-reckoning error-resetting \cite{wang2012no,abdelnassersemanticslam}.  Finally, fine-grained tracking of passengers' activities and interaction with the different detected semantics opens the door for indoor analytics, which is of great business value.\\
In this paper, we present \sys{} as a crowdsensing system that leverages the ubiquitous sensors available in commodity cell-phones to automatically  enrich transit stations  floorplans with  different semantics. The core idea is that passengers  perform many activities (e.g.,  crossing an entrance gate) that show identifiable signatures on the phone sensors. \sys{}  aims  to recognize  these high level activities and therefrom   discover their uniquely  associated  semantics. Therefore, starting from an \textbf{unlabeled general} floorplan of a transit station, \sys{} estimates the location of different semantics and tags their locations on the map  accordingly  to generate a detailed  floorplan  (Figure \ref{eg}). \\
Translating this basic idea into a deployable system, however, involves addressing a number of challenges: First, identifying semantics signatures is based on the phone sensors during passengers' activities;  which are prone to  human behavior artifacts. Second, current indoor localization technologies may require infrastructure support or prior calibration; and all have an average localization error in the range of few meters \cite{youssef2015towards}. This can place the passenger in  a  location on the floorplan that is  far from the actual one, affecting the accuracy of semantics' location estimation. Finally, the system needs to be optimized for energy to avoid significant battery drainage.\\
To cope with these challenges, \sys{}
draws on a classifier-based approach based on the multi-modal sensors
features to recognize passengers activities and thereby identify
their associated semantics to address the first challenge. For the
second challenge, \sys{} relies on a DBSCAN clustering algorithm to cluster the correct crowdsensed samples of the same
semantic to estimate its location and thus outlier locations are removed.
To save energy, \sys{} employs sensors with low
energy footprint (i.e., inertial sensors) and the energy hungry sensor employed (i.e., sound) is turned
on only when needed.\\
 Implementation of \sys{} over different Android phones shows that it can detect the fine-grained stations semantics accurately  with 7.7\% false positive rate   and 7.5\% false negative rate  on average. In addition, it can estimate  locations of the detected semantics accurately,  achieving an error of 2.5m  on average using as few as 40 samples of  each semantic. This comes with low energy consumption of 41 Joule on  average for typical traces. We believe this could be a promising  and a potential candidate for the real-world.\\
In summary, our contributions are three-fold:
\begin{itemize}  [noitemsep,topsep=0pt]
\item We present the \sys{}  system to  automatically  and unobtrusively crowdsense and identify transit stations semantics (e.g., ticket and drink vending machines, entrance gates, lockers, waiting (sitting) areas, restrooms, platforms and cars' waiting lines, escalators, elevators and stairs) from  the available  sensors readings with minimal energy consumption.
\item We provide a  framework for  extracting  the features  used to recognize high level user contexts (e.g., buying a ticket) from a sequence of temporal  and spatial low level user states (e.g., walking, standing, etc)
based on the phone sensors.
\item  We   implement \sys{} on Android phones,  collect  real data by 16 participants, and  evaluate its accuracy, generalizability, robustness  and energy-efficiency at eight different railway stations in Japan.

\end{itemize}
The rest of the paper is organized as follows: Section \ref{sec:ove} presents the system overview. We give the details of identifying  station's semantics  from phone sensors  in sections \ref{sec:level}  and  \ref{sec:act}. Section \ref{sec:eval} provides the  evaluation of \sys{}. Section  \ref{sec:dis} discusses the system limitations and possibilities for enhancement. Finally, sections  \ref{sec:related} and \ref{sec:con}  discuss related work and  conclude  the paper respectively.
\section{The TransitLabel System}
\label{sec:ove}
\begin{figure}[t!]
\centering
\includegraphics[height=7cm]{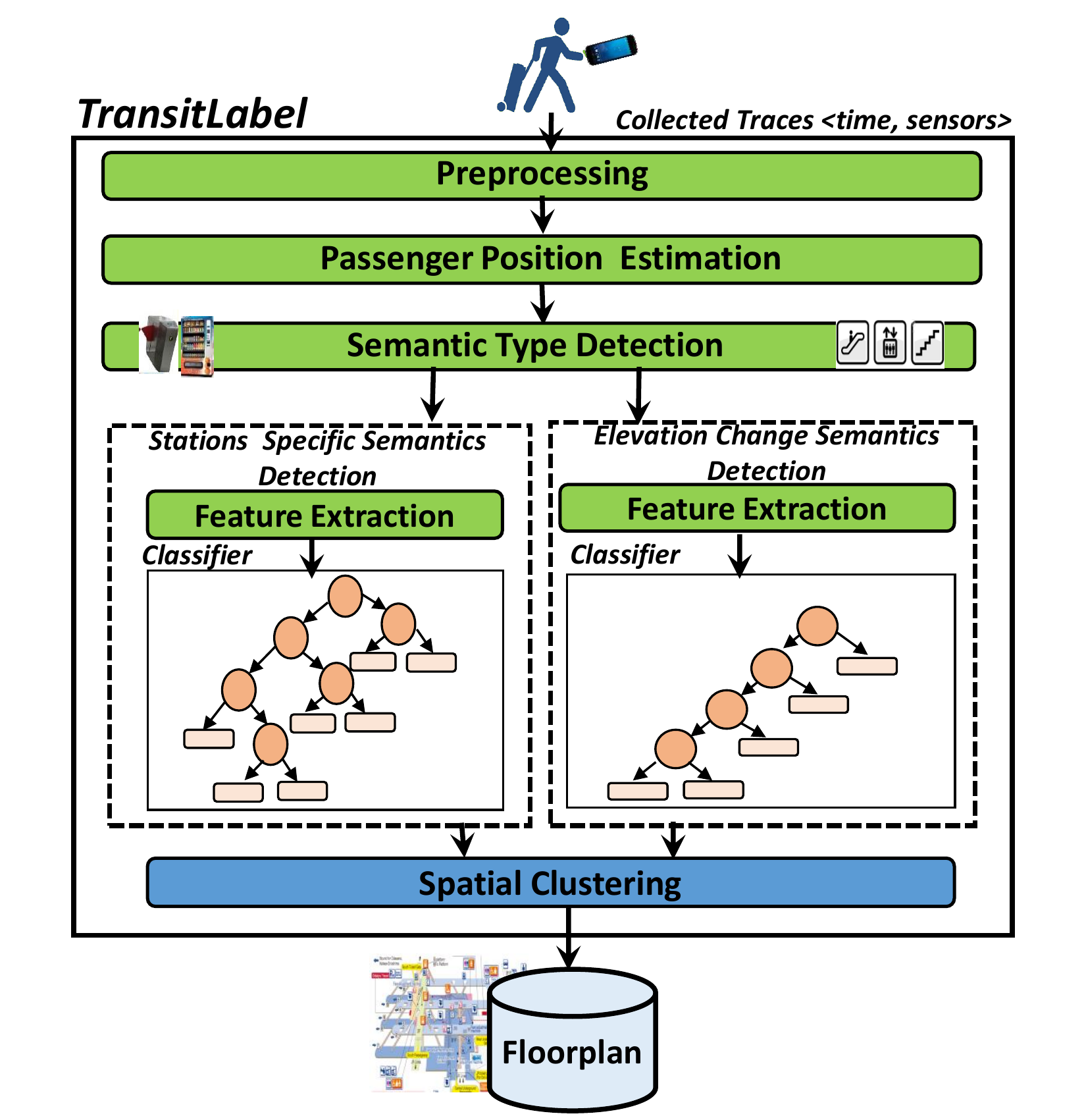}
\caption{The \sys{} system architecture.}
\label{arch}
\end{figure}

Figure \ref{arch} shows the \sys{} system architecture. \sys{} is based on a crowdsensing approach, where cell phones carried by users submit their data to the \sys{} service running in the cloud. The data is first preprocessed  to reduce the noise. Then, semantics are classified to separate the elevation change semantics (elevators, escalators, and stairs) from other railway stations exclusive semantics (ticket vending machines, entrance gates, etc). \sys{} has two core components: one for extracting elevation change semantics and the other for extracting  other stations exclusive semantics. \sys{} takes a classifier approach
to detect  different  semantics based on the extracted features from the collected sensor traces. In the rest of this section, we give an overview of the system architecture leaving the details for the semantics detection to sections \ref{sec:level} and \ref{sec:act}.
\subsection{Traces Collection}
 The system collects time-stamped  sensor measurements including the available inertial sensors (accelerometer, gyroscope and magnetometer), barometer as well as the sound sensor (i.e., the microphone).
 Inertial sensors have a low-cost energy profile and they are already running all the time during the standard phone operation  to detect phone orientation changes. Therefore, they consume zero extra energy. On the other hand,  we use  an adaptive sensor scheduling scheme called  \textit{triggered sensing} \cite{mohan2008nericell} to reduce the sound sensor energy consumption.  The key idea is that  sensors that are inexpensive in energy consumption (e.g., accelerometer) are used to trigger the operation of more expensive sensors (e.g., sound). Specifically, \sys{} activates audio recordings  only as soon as the  passenger becomes stationary  for a considerable time (4 seconds) and  suspends it once she resumes walking. The intuition  is that passengers traces inside railway stations are  dominated by walking  periods  and  they  pause only to perform an activity (e.g., buying a ticket or a drink) which has a considerable  stationarity
time (more than 15 seconds). If the user does not resume walking after a certain time (60 seconds), the audio recordings will be suspended to save energy.
The  collected audio recordings during activities  are used as a tie-breaker  when other sensors (e.g., inertial sensor)  fail to recognize certain activities. \\
Given the privacy implications of turning on the microphone, \sys{} gives users full control over their own sensed data by means of a personalized privacy configuration. \sys{} has different modes of operations (full sensor collection, privacy insensitive data only) that tailor the amount of data collected based on the user's preferences. In addition, according to a recent study\cite{chon2013understanding}, inertial sensors are enabled by most users and even the privacy-sensitive sensors (i.e., microphone) are enabled by about 78\% of users. Finally, we \textbf{process audio data locally}  on the user's device to further enhance the user privacy.
 \subsection{Preprocessing}
 This module is responsible for preprocessing the raw  sensor measurements to reduce the effect of (a) phone orientation changes and (b) noise, e.g.,  small  direction changes while moving. To handle the former, we transform the sensor readings from the mobile coordinate system to the world coordinate system leveraging the inertial sensors \cite{mohssen2014s}. To address the latter, we apply a low-pass filter to raw sensors data using local weighted regression  to smooth the data \cite{cleveland1988locally}. To  filter out the noise in  the employed frequency bands (350Hz and 3kHz) in  audio recordings,  the standard sliding window averaging technique, with a window of 32 samples, is used.
\subsection{Passengers' Position Estimation}
\sys{} needs accurate passengers' locations  during activities  to  estimate their uniquely associated  semantics positions. To achieve this, \sys{}  employs the  dead-reckoning  technique  to track  the  passenger's location  starting from a reference point (e.g., the station entrance). We  employ the step detection algorithm in \cite{alzantot2012uptime} that takes into account the different users' profiles and gaits and apply them to the acceleration signal to detect the user steps. The user heading is estimated by incorporating the algorithm in \cite{abdelnassersemanticslam,wang2012no} which leverages the correlation between compass and gyroscope to compensate gyroscope drift and compass interference errors to accurately estimate the user orientation. The estimated displacement and heading are fused to localize the user. However, the displacement error of dead reckoning is unbounded making it infeasible for indoor tracking. To alleviate this problem, \sys{} incorporates  the  idea  of SemanticSLAM  \cite{abdelnassersemanticslam,wang2012no}  by leveraging \textbf{amble and unique} physical  points in the stations  (i.e., semantics) to reset the accumulated error.  Since dead-reckoning provides a rough  location to the phone, it is also possible to roughly localize the semantics based on when the phone senses them. Now, since the  floorplan is known,  we can  estimate  the locations of all semantics  in a crowd-sensing approach (as discussed later) by combining the rough estimates (i.e., the dead-reckoned positions) from multiple passengers' phones. These semantics, once detected  based on their unique sensor signatures, can then be used to improve dead-reckoning of subsequent passengers, which in turn can refine the semantic locations. This recursive dependence between estimating the semantic  location and the user location is similar to the Simultaneous Localization And Mapping (SLAM) framework.
\subsection{Semantic Type Detection}
\sys{} is designed to detect various stations semantics based on their  unique usage patterns. This  module separates between the two major types of semantics: elevation change semantics (elevators, stairs and escalators) and  stations specific semantics (e.g., ticket vending machines, entrance gates, etc). The usage of elevation change  semantics  involves a noticeable  change in the  passenger's level (i.e., height) which is absent in  other  semantics  (Figure \ref{level}). To separate them, we draw on the maximum  difference among the relative  barometer readings (i.e.,  pressure)  in consecutive  overlapping windows. The  employed window size, 10 seconds, is small  enough so that  barometer readings are not affected by the environmental changes \cite{muralidharan2014barometric}. The intuition is that a change in pressure means a change in height which in turn means that the passenger is using  one of the elevation change semantics. Moreover, the sign of the pressure difference indicates the direction of motion (up or down) which  is  useful for other purposes (e.g., detecting escalators direction). Evaluation of over 250 traces shows that the semantic type detection can achieve 0\% false positive and negative rates. Later, the major two classes of  semantics are further classified  to  their more fine-grained  semantics.
 \begin{figure*}[!t]
        \centering
        \begin{subfigure}[b]{0.32\textwidth}
                \includegraphics[width=\textwidth,height=4.5cm]{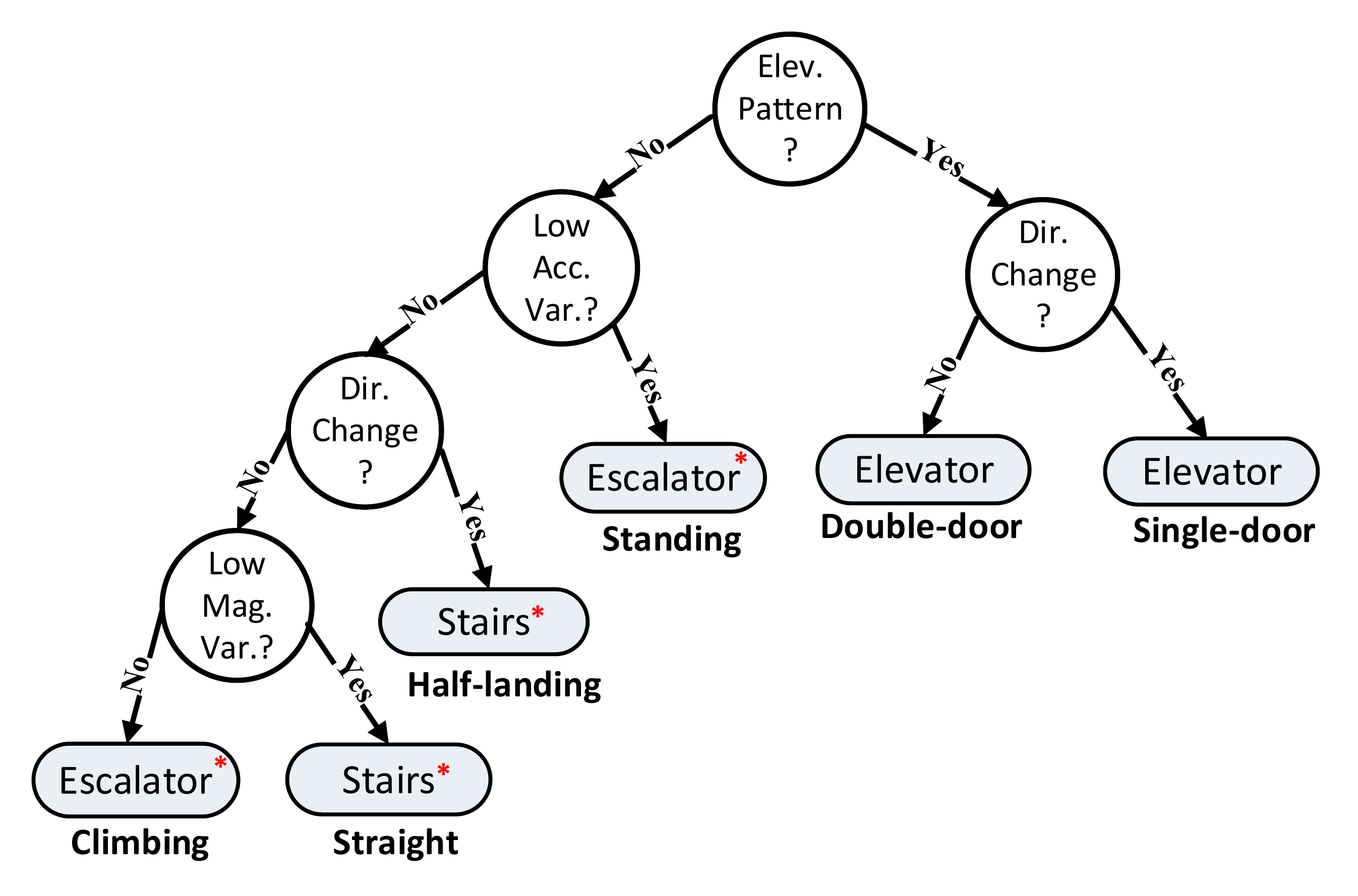}
           \caption{Elevation change semantics. Star marked semantics are further classified as floor or level change semantics leading to  different floors or different levels within the same floor respectively.}
           \label{transport}
        \end{subfigure}%
        \begin{subfigure}[b]{0.72\textwidth}
                \includegraphics[width=\textwidth,height=5cm]{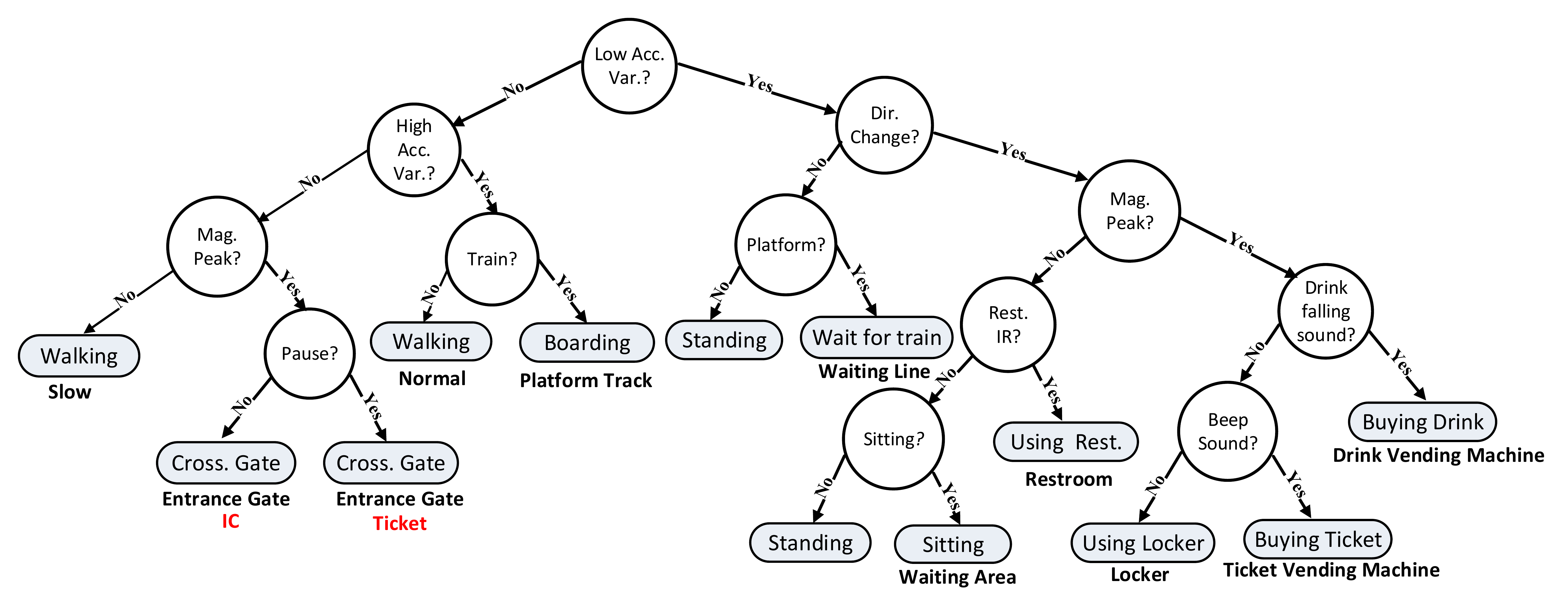}
              \caption{Stations specific semantics.}
              \label{other}
        \end{subfigure}

        \caption{A decision tree classifier for detecting  different types of  semantics.}
\end{figure*}
\subsection{Feature Extraction}
In this section, we present the basic features  used to identify the different semantics based on the data collected from the passengers' phone sensors. For instance, magnetic peak  is  a key feature  to recognize many activities  that involve direct interaction with electronic machines (e.g., vending machines). To extract it,  we first apply a stream-based event detection algorithm to identify significant changes in the \textbf{magnitude} of the magnetic field. Once a significant change (a \textit{10uT}  increase  in a window of \textit{50} samples\footnote{We experimented with different values of  thresholds and  selected values that  are robust to changes in the platforms/stations as confirmed by our experiments.}) has been observed, we mark the corresponding time instant as the starting boundary of the peak area. We buffer subsequent measurements until a significant decrease in the magnitude of the magnetic field  is observed. Once the starting and ending boundaries have been identified, we extract two  features that characterize the peak area: the peak duration and  strength.  Moreover,  some activities (e.g., buying tickets) are characterized by  a sudden change in the user direction (i.e., surge in gyroscope readings) during or directly  after the activity period. To detect this sudden change, we used the \textit{approximate derivative} method: The derivative of sensor  values within a time window are compared against a predetermined threshold (\textit{75\degree}  in a window of \textit{60} samples) to detect the  surge in  sensor values. Finally,  the variance  of  the acceleration   is used  to discriminate various passenger motion types (stationary, slow walking and normal walking) which contributes to the identification of  many higher  level passenger activities (e.g, buying a ticket). The  acceleration variance values  of \textit{1.8} and \textit{7} for a window  of \textit{200} samples  are used as thresholds to separate  stationary,  slow walking and normal walking motion patterns respectively.
\begin{figure*}[!t]
\noindent\begin{minipage}[t]{0.23\linewidth}
\includegraphics[height=3.5cm,width=1.1\linewidth]{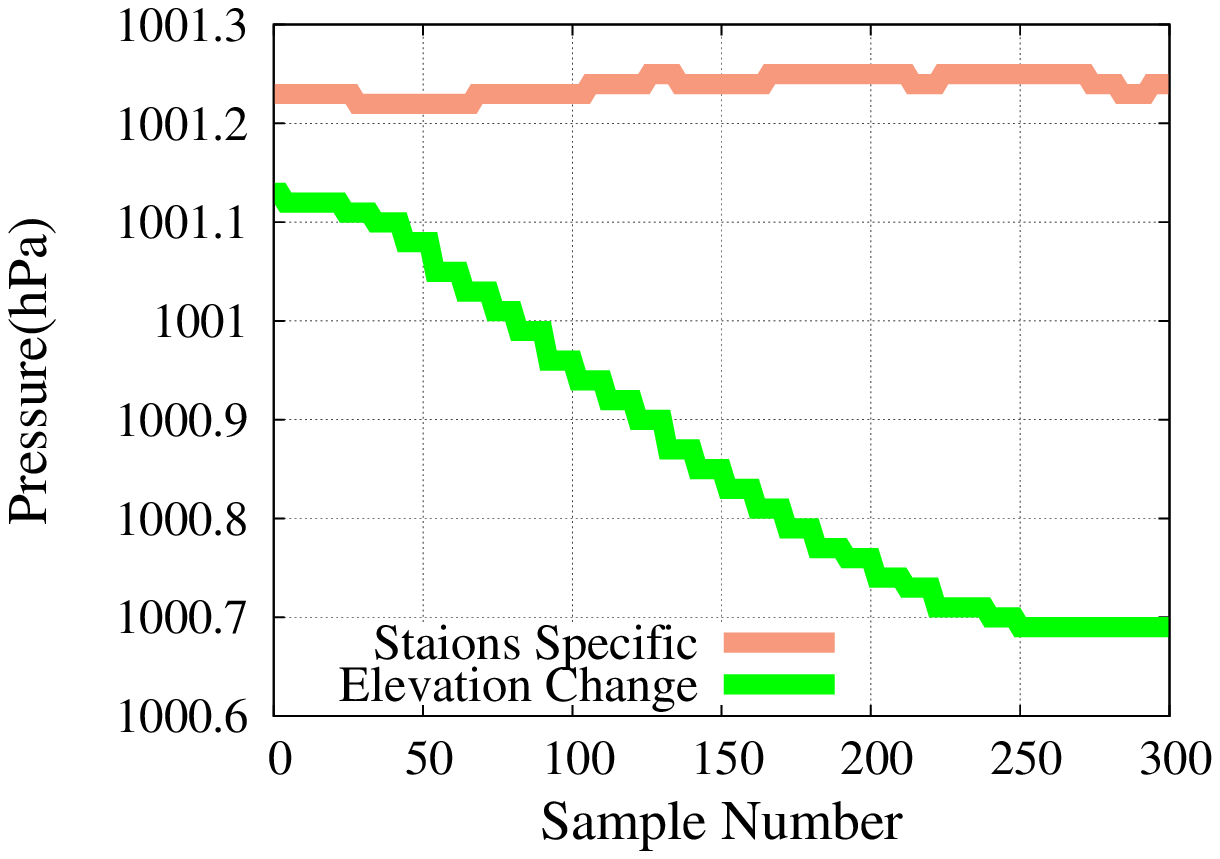}
\caption{Comparing barometer readings of  the usage of elevation change semantics (e.g., elevator) against  station specific machines (e.g., ticket gate).}
\label{level}
\end{minipage}
\hfill
\begin{minipage}[t]{0.24\linewidth}
  \includegraphics[width=1\textwidth,height=3.3cm]{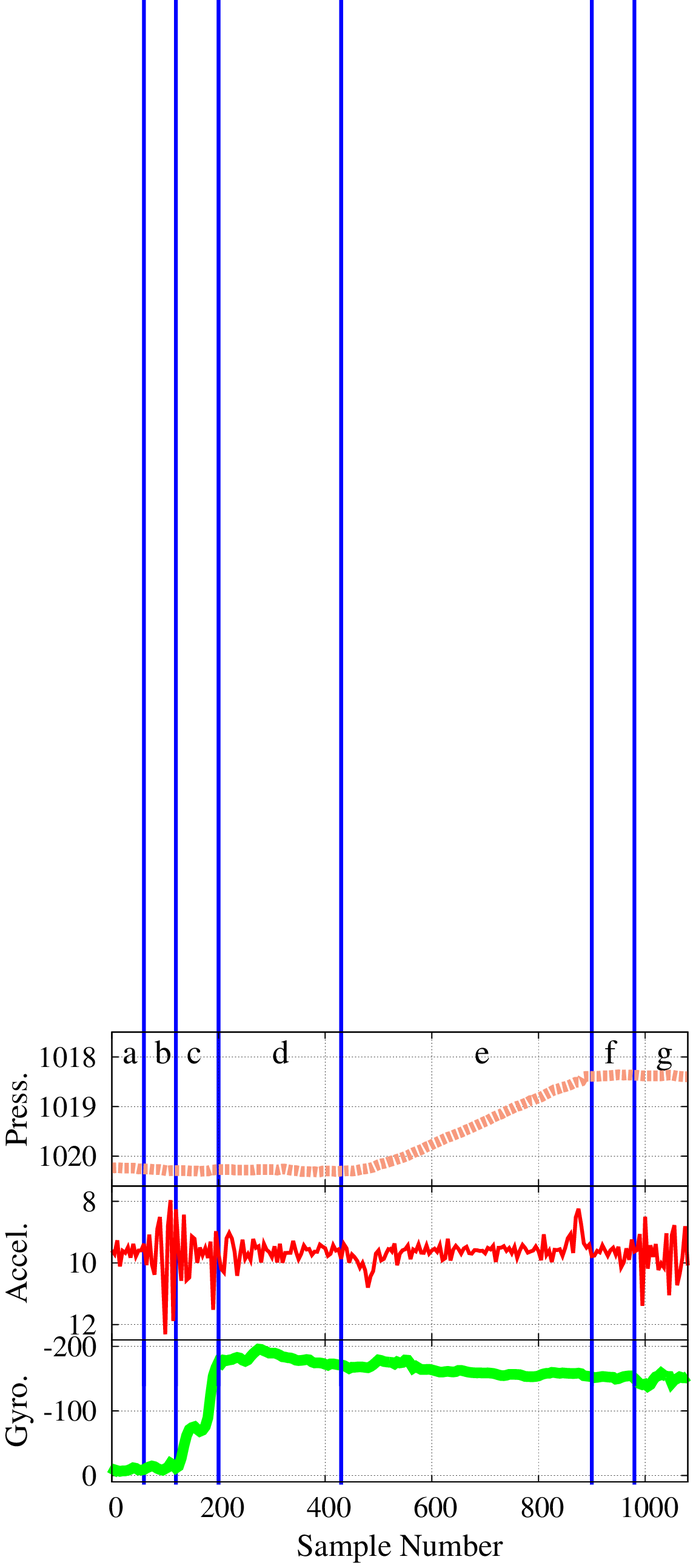}
  \captionof{figure}{Elevator  usage  pattern: (a) waiting for it, (b) walking into it, (c) direction change, (d) stationary, (e) going \textbf{up}, (f) stopping,  (g) walking out.}
   \label{elevator}
\end{minipage}
\hfill
\begin{minipage}[t]{0.24\linewidth}
\includegraphics[width=1\textwidth,height=3.5cm]{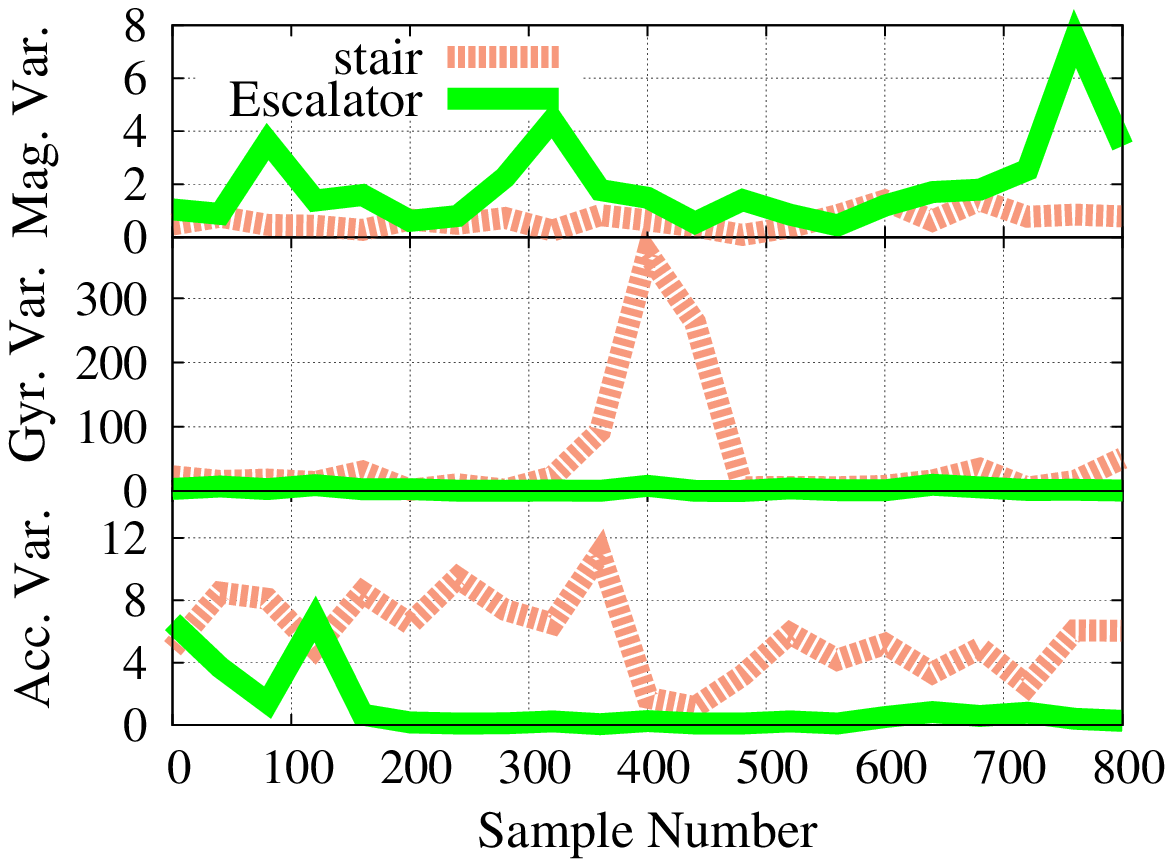}
\captionof{figure}{The sensor patterns that compare  climbing up a  half-landing stairs  against standing on a moving up escalator.}
\label{esc}
\end{minipage}
\hfill
\begin{minipage}[t]{0.24\linewidth}
\centering
    \includegraphics[height=3.5cm]{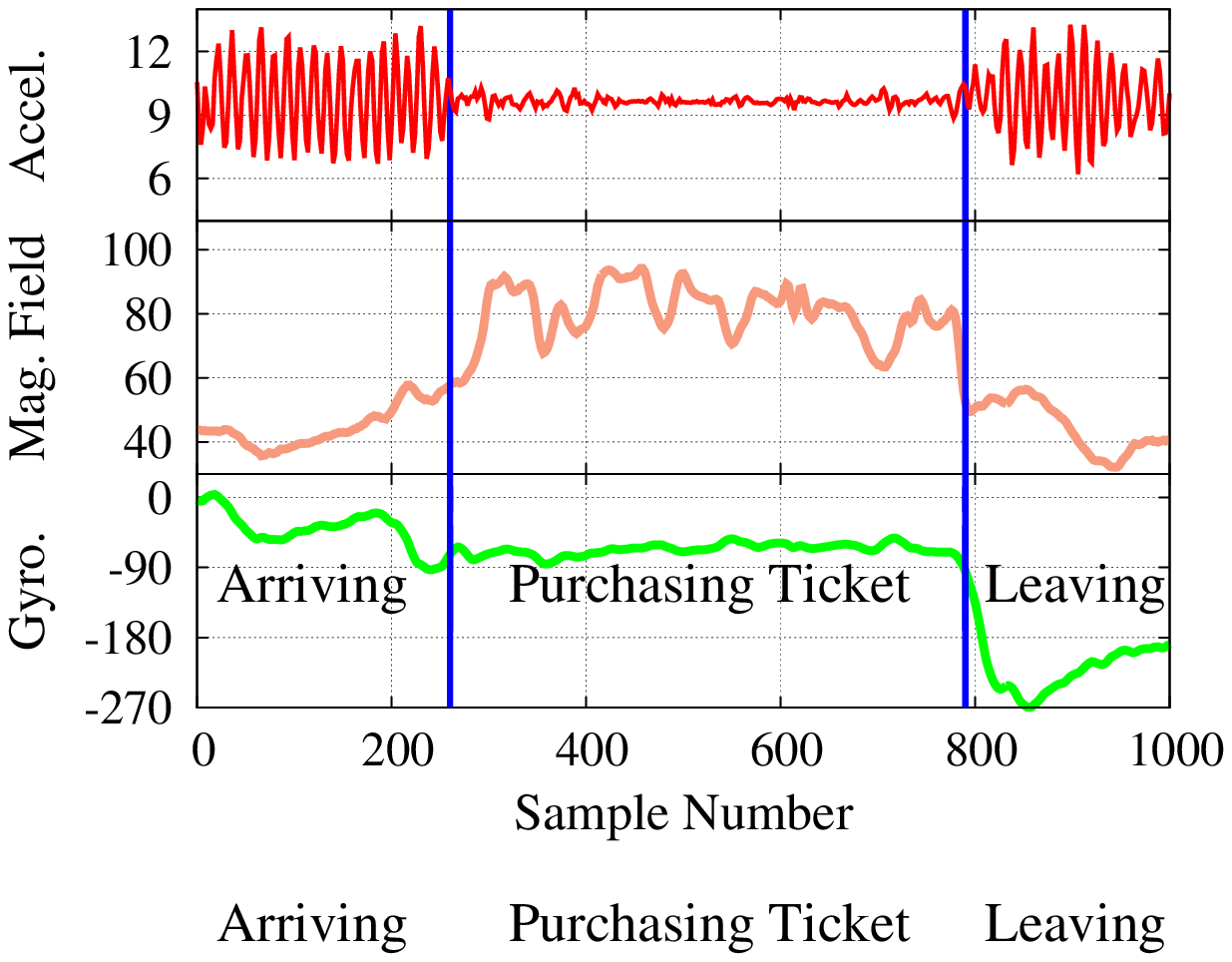}
    \caption{The acceleration, ambient magnetic field and gyroscope readings while using a coin operated machine.}
\label{fare}
\end{minipage}
\end{figure*}
\subsection{Station Semantics Extraction}
 As many semantics share some sensors patterns while having different patterns on other sensors, the hierarchical classification is an intuitive solution. Therefor, to identify   semantics, \sys{} relies on a tree-based classifier  as it is easy to  understand and to generate its rules. This classifier  decomposes the task hierarchically into subtasks, proceeding from a coarse-grained classification (shared patterns)  towards the distinction of fine-grained  semantics (distinctive patterns) as detailed in sections \ref{sec:level}  and  \ref{sec:act}.
\subsection{Semantic  Location Estimation}
Whenever a  semantic  is detected by the semantic detection modules discussed later, \sys{} needs to determine whether it is a new instance of  a station  semantic (i.e., not discovered before) or not as well as determine its location.\\
 To do this, \sys{} applies spatial clustering for each type of the extracted semantics. It uses the density-based clustering algorithm (DBSCAN \cite{ester1996density}) which has a number of advantages as the number of clusters is not required before carrying out clustering; the detected clusters can be represented in an arbitrary shape; and outliers can be detected. The DBSCAN is applied to all  samples of each  discovered  semantic to cluster all samples that are adjacent to each other in the spatial space. The parameter \textit{Eps}  specifies  the radius of each cluster controlling  the maximum distance among samples on the same cluster. After clusters are formed, the locations of the newly discovered semantics are estimated as  the weighted mean of the points inside their clusters. We weight the different locations based on their location accuracy reported by the localization approach. Specifically,  in our position estimation  approach, the longer the user trace from the last resetting point, the higher the error in the trace. Therefore, shorter traces have better accuracy. Based on the law of large numbers, the weighted average of independent noisy samples should converge to the actual location of the semantic. When a new semantic is identified, if there is an already discovered  semantic within its neighborhood, we add it to the cluster and update its location. Otherwise, a new cluster is created to represent the new  semantic. To reduce outliers, a semantic is not physically added to the floorplan until the cluster size reaches a certain threshold which is specified by the \textit{Minpts} parameter (the minimum number of points that can form  a cluster) of  the DBSCAN algorithm. The DBSCAN parameters \textit{Minpts} and \textit{Eps} are selected empirically for each semantic type depending on its available number of  samples,  its  physical dimension, and the average inter-distance among its physical  instances in the real-world stations indoor maps. We do not state the DBSCAN parameters values  for  each semantic due to space constraints.
\subsection{Practical Considerations}
Sensor specifications are different from one phone manufacturer to another, which leads to different sensor readings for the same activity. To address this issue, \sys{} applies a number of techniques including use of offset-independent features (e.g., variance), orientation independent features (e.g., magnitude of acceleration and  magnetic field) and combining a number of features for detecting the same semantic. In addition, we experimented  with various  thresholds and select those leading to \textit{high detection accuracy} with low false positive/negative rates while being  robust to different users and stations. This is confirmed by experiment performed in the evaluation section.\\
\sys{} does not also require real-time sensor data collection (i.e., it works offline); it can store the different sensor measurements and opportunistically upload them to the cloud for processing; allowing it to save both communication energy and cost. This
is outside the scope of this paper.
\section{Elevation Change  Semantics Detection}
\label{sec:level}
To classify elevation change semantics \footnote{We provide a thorough comparison with other related  approaches in the related work section.} into their fine-grained classes (elevators, escalators and staircases), we  apply  a decision tree classifier to the extracted  features from passengers' phone sensors based on our observations of semantics usage scenarios and their physical structures (Figure \ref{transport}).
\begin{figure*}[!t]
\noindent\begin{minipage}[t]{0.48\linewidth}
  \centering

          \begin{subfigure}[b]{0.32\textwidth}
                \includegraphics[width=\textwidth]{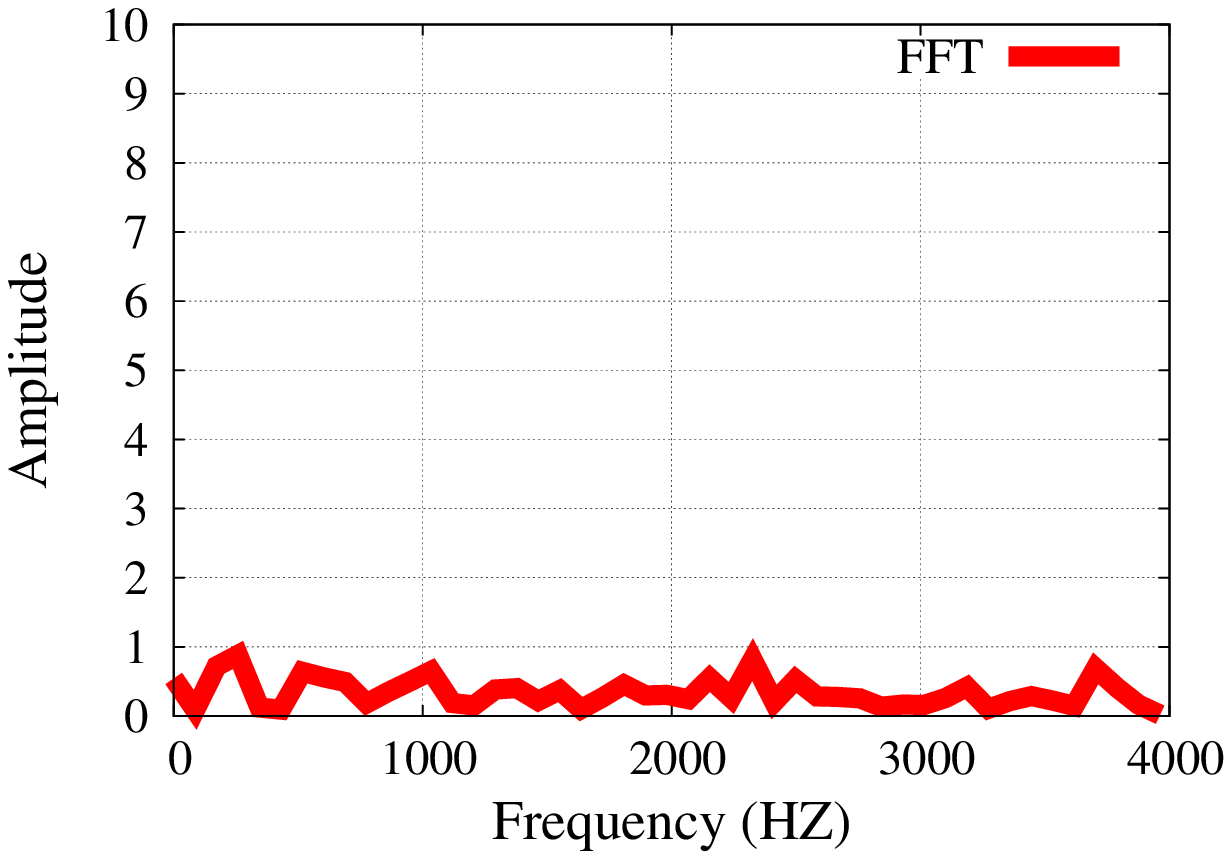}
           \caption{Backgr. noise.}
        \end{subfigure}%
        \begin{subfigure}[b]{0.32\textwidth}
                \includegraphics[width=\textwidth]{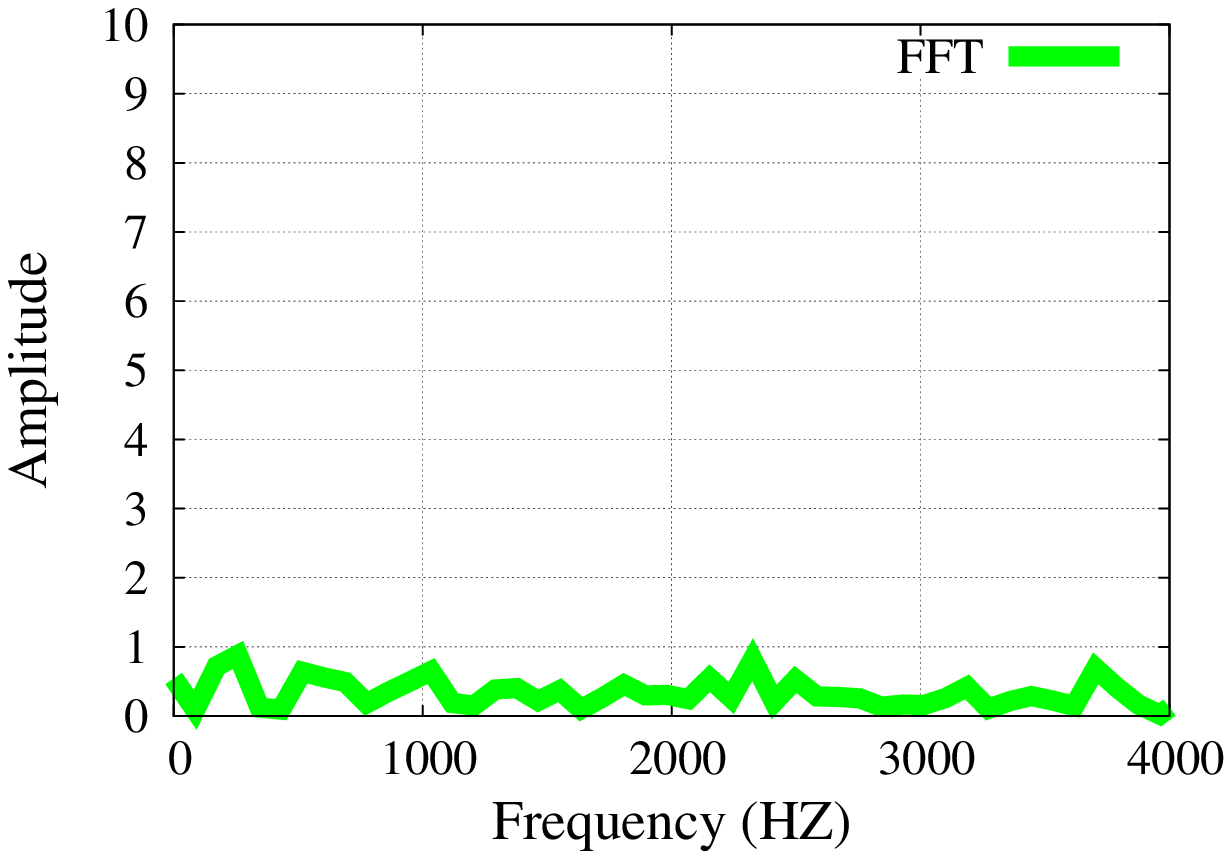}
           \caption{Coin insertion.}
        \end{subfigure}%
        \begin{subfigure}[b]{0.32\textwidth}
                \includegraphics[width=\textwidth]{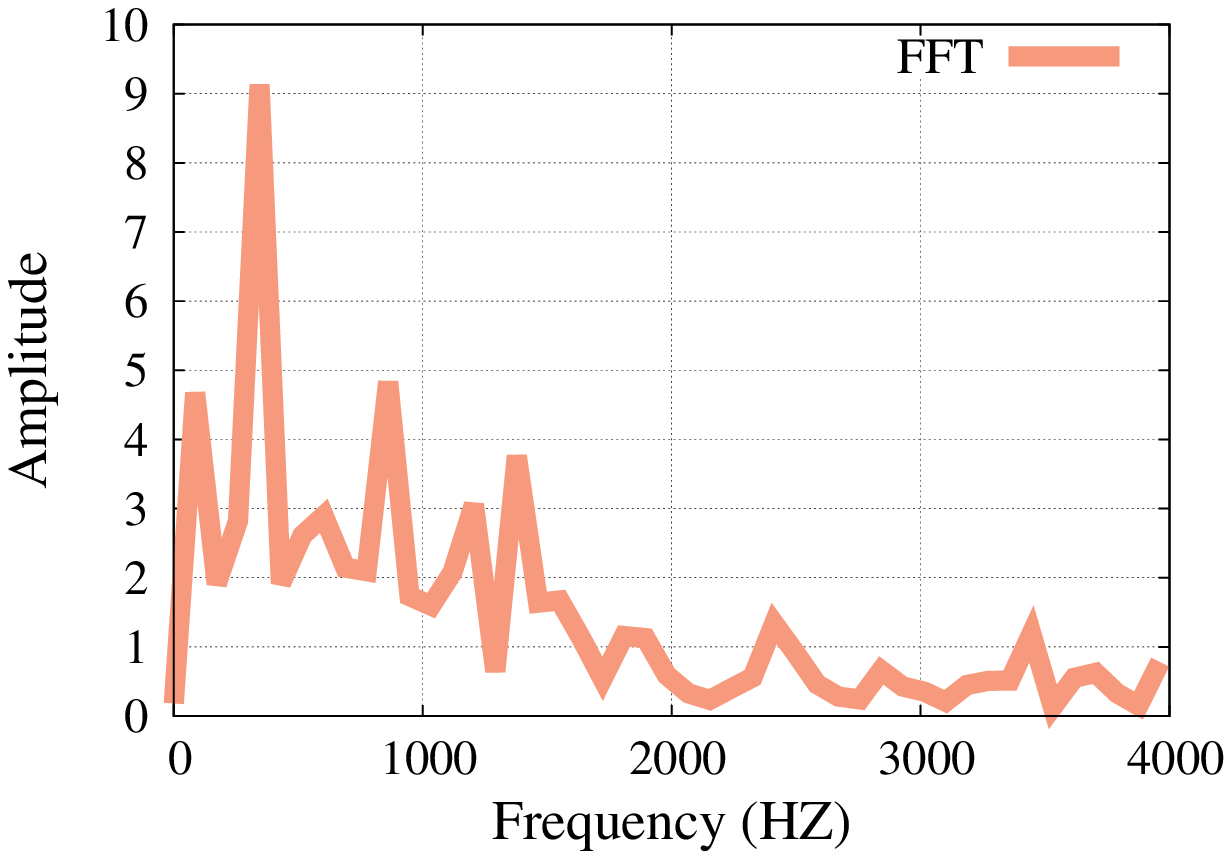}
              \caption{Drink falling.}
        \end{subfigure}

            \begin{subfigure}[b]{0.96\textwidth}
                \includegraphics[width=\textwidth,height=2.8cm]{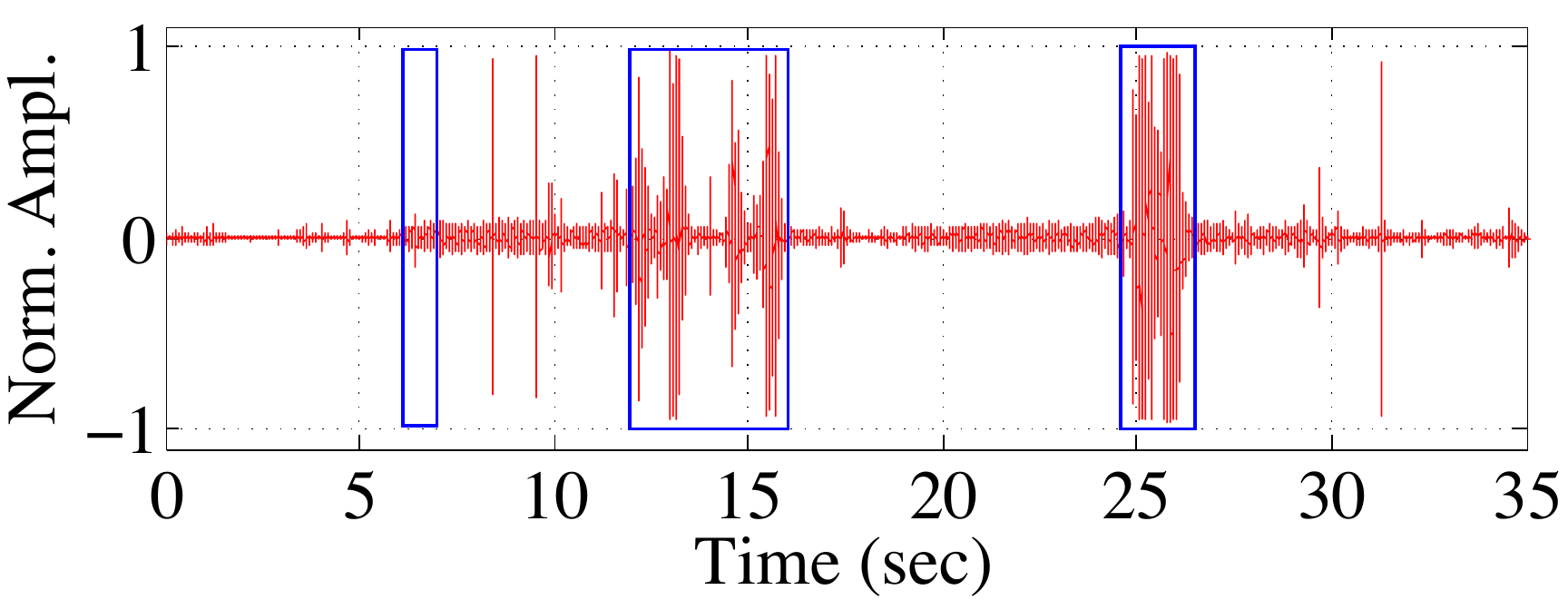}
              \caption{The original audio signal in time domain}
        \end{subfigure}

        \caption{ A time-domain sample of an audio signal depicting the usage of a drink vending machine is shown in (d) where three highlighted audio signal  (bounded by  blue boxes)  corresponding to the background noise, the coin insertion sound, and the drink falling sound respectively. Figures (a), (b), (c) depict the frequency domain of these three audio signals respectively.}
\label{fig:ticketvsdrink}
\end{minipage}
\hfill
\begin{minipage}[t]{0.48\linewidth}
 \centering

          \begin{subfigure}[b]{0.4\textwidth}
                \includegraphics[width=\textwidth,height=1.8cm]{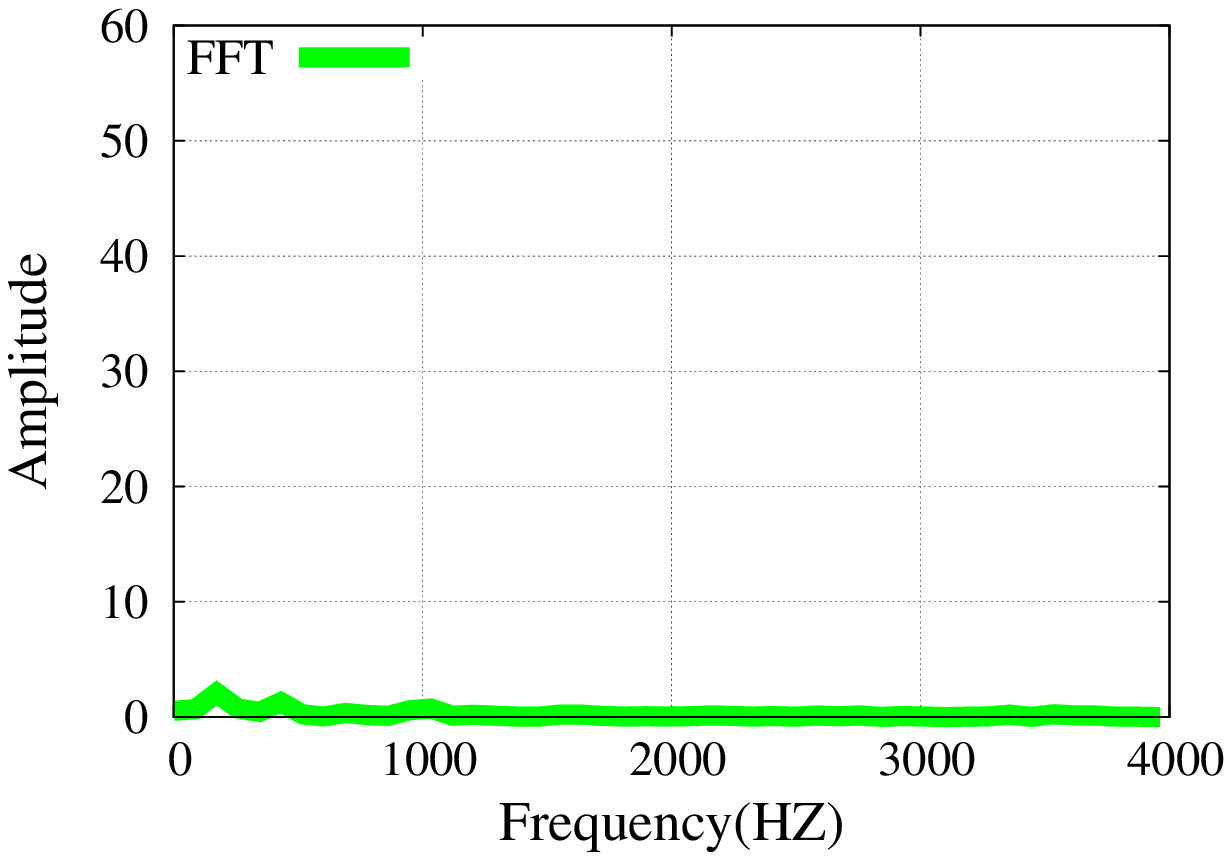}
           \caption{Background noise.}
        \end{subfigure}
        \begin{subfigure}[b]{0.40\textwidth}
                \includegraphics[width=\textwidth,height=1.8cm]{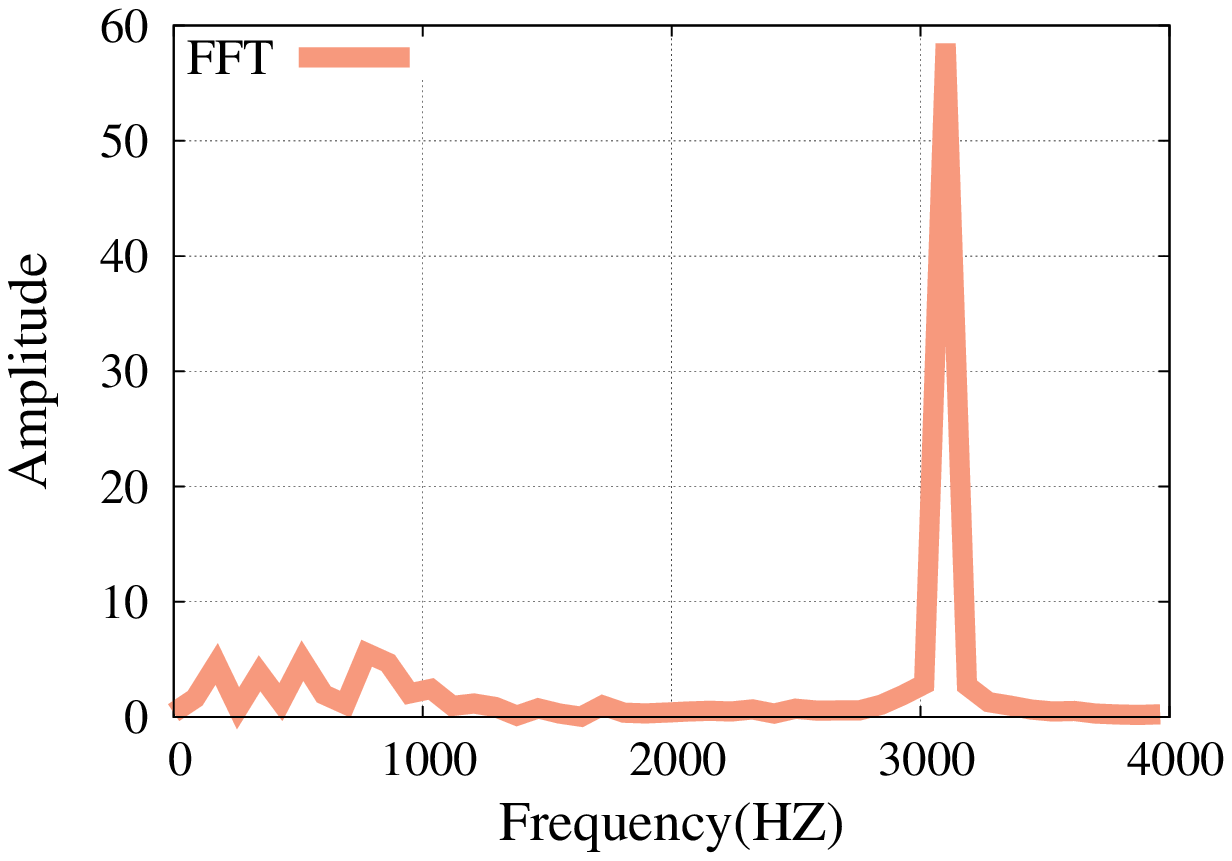}
           \caption{Beep signal.}
        \end{subfigure}

            \begin{subfigure}[b]{0.92\textwidth}
                \includegraphics[width=\textwidth,height=2.8cm]{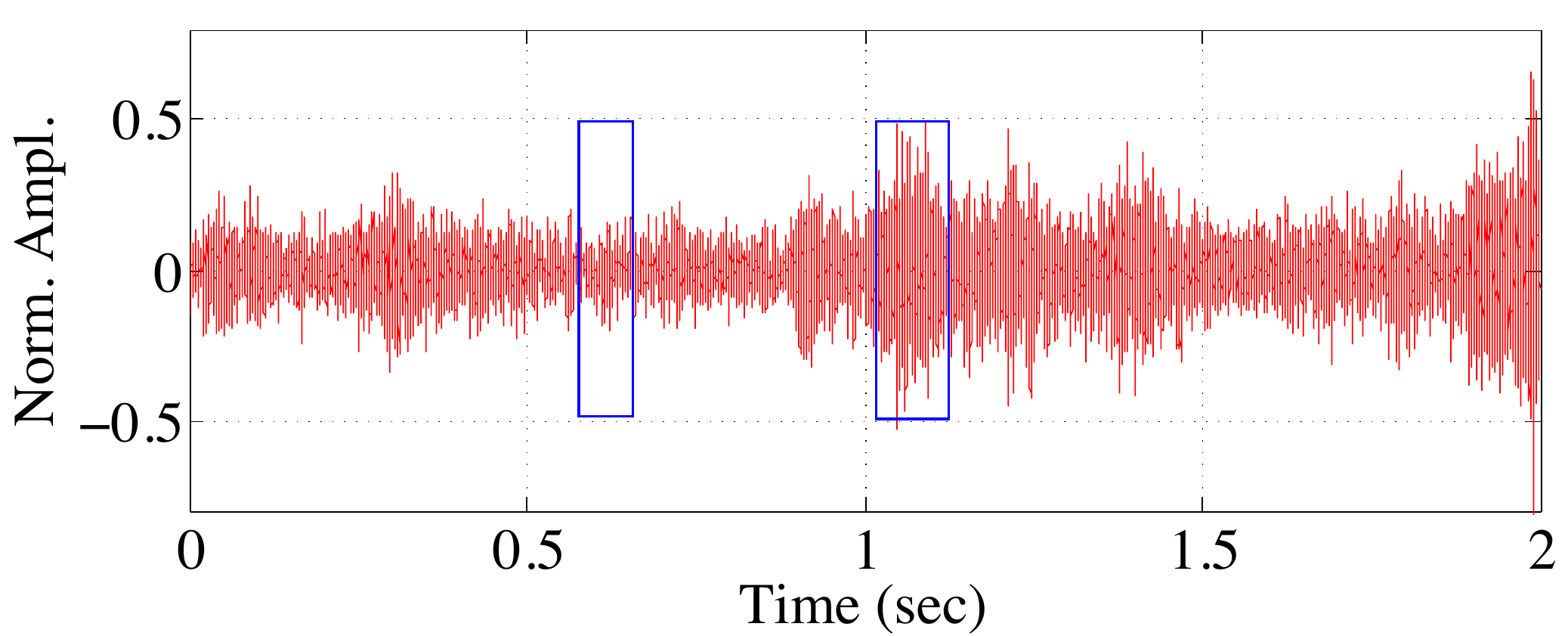}
              \caption{The original audio signal in time domain}
                  \end{subfigure}

        \caption{ A sample of audio signal  in the time domain depicting the usage of a ticket  vending machine is shown in (c) where two different audio signal are highlighted (bounded by  blue boxes) corresponding to the background noise  and the  beep audio signal respectively. Figures (a) and (b)  depict  the frequency domain of these  two cropped  signals respectively.}
\label{fig:ticketvslocker}

\end{minipage}
\end{figure*}

\noindent  \textbf{Elevators:} \\ We  begin by recognizing elevators  as it  is straightforward to distinguish their unique pattern. The  typical usage  scenario of an elevator consists of a normal working period, waiting for the elevator, walking into the elevator,  changing direction to face the exit door, standing for a while, followed by a  level  change  when it starts to move  (Figure \ref{elevator}). This behavior is reflected to a  unique   pattern   that consists of  sequence of states:  walking, stationary, stepping,  direction change, level change,  and Accelerate-Stationary-Decelerate emerging from the start and stop of the elevator. A set of features are  extracted from accelerometer,  gyroscope and barometer readings and fed to a Finite State Machine (FSM) to detect  this multi-modal pattern. Starting from  the initial state which represent the user waiting for an elevator (state (a) in Figure \ref{elevator}), the  transitions  through  all subsequent states (from (b) to (g)) have to occur to announce that  a \textbf{single-door elevator} have been detected. However,  for elevators having two doors, users do not need to turn around. Nevertheless, \textbf{two-doors elevators} can be recognized by the same FSM while skipping the direction change transition state.\\
\textbf{Escalators (Standing):} \\ The  acceleration variance is used to decide whether a user is standing on an escalator or not. The intuition is that when a user keeps standing while carried by a  moving staircase, the acceleration variance  remains  small compared  to climbing stairs or escalators, which generates a  high  acceleration variance  due to  the vertical motion of the user. Conversely, if  the acceleration variance  is high, we cannot verify whether the user is climbing  a stair or an escalator. \\
\textbf{Half-landing  Stairs (Climbing):} \\ Half landing staircases  have a  turn in the  middle forcing  users to change their direction while straight stairs  and escalators do not have any turns. Thus,  if there is   a surge in gyroscope  readings  (from the direction change) that  took place  in the  middle of the elevation change period, it is  affirmative that the user is climbing a  half-landing stairs.  Otherwise, if there is no direction change,  we cannot  verify  whether a user is climbing a straight stair or climbing an escalator.\\
\textbf{Escalators (Climbing):}  \\ The magnetic field variance, due to the escalator constant speed  motors, can be used to reliably differentiate between climbing a straight stair and climbing an escalator (Figure \ref{esc}). The  value  \textit{100}, in a window of \textit{200} samples,  is used  as the threshold of the variance of magnetic field.\\
\textbf{Straight Stairs (Climbing):} \\After separating other elevation change semantics, the remaining  samples are classified as straight stairs. \\
\textbf{Level Change or Floor Change:} \\ Many stations are multi-floor buildings with a typical floor height  between 3.0 to 6.0 meters. The majority of elevation change semantics installed in railway stations move passengers from one floor to another (Floor change semantics). However,  there exist  some low height stairs and escalators which move  passenger from level to another within the same floor (level change semantics). To classify the type of escalators and stairs (marked by red stars in Figure  \ref{transport}),  we rely on the magnitude of pressure difference during the elevation change  period.  Given that 1.0 meter height  change corresponds to 0.12 hPa change in pressure, the pressure difference of  \textit{0.3} hPa  is used as a threshold to separate  level change escalators and stairs from floor change ones.
\section{Station Specific Semantics Detection}
\label{sec:act}
Stations are rich  with many  exclusive semantics  like ticket vending machines, entrance gates, drink  vending  machines, platforms, cars' waiting lines, lockers, restrooms,  and waiting (sitting) areas. Based on our observations, these semantics force users to behave in predictable ways  which are translated to unique sensor signatures that can be mined to identify them. For instance, a  passenger crossing an entrance gate has to slow down her walking speed  until she pauses to drop the ticket into the  gate machine and  then steps forward to grab it.  Meanwhile   her phone is experiencing a magnetic field distortion emanating  from the  gate machine electronics. \sys{}  draws on a  decision tree classifier   to recognize   different passengers'  activities  (Figure \ref{other}). The root of the  decision tree separates  activities  into  two  main  classes. The right branch of the tree  comprises activities  that  require  the passenger to be   stationary during the service time (ticket and drink vending machines, lockers, restrooms, waiting (sitting) areas, etc). On contrast, the left branch  comprises activities that do not force  passengers to pause (e.g., ticket gates, etc). In the balance of this section, we give the details of the classifier features
that can differentiate the different station specifics semantics (coin operated machines, entrance gates, platforms track, sitting areas, restrooms).
\subsection{Coin Operated Machines}
 Nowadays, ticket vending machines are found in every station and drink vending machines exist  in many  transits to dispense items (e.g. beverages, etc) to customers automatically. In addition, coin operated lockers  are widely installed  in railway stations  to allow  passengers to leave their baggages for several hours respectively to visit the surrounding area freely (especially in major stations in the downtown areas).\\
To identify these machines,  we observed that their  typical usage  traces consist of normal  walking  to the machine, followed by standing  in front of it, inserting currency, beginning the service (choosing a drink or  the ticket type in case of drink and ticket vending machines respectively or opening  the locker door in case of lockers),  finishing the service (grabbing the ticket or the drink in case of ticket and drink vending machines respectively or putting  luggage  into  the drawer and locking it), and finally  walking away (Figure \ref{fare}).  This usage scenario is translated to the following unique patterns on the sensors. First, the user is stationary during the machine usage. Second,  there is a fluctuation in  the magnetic field readings as soon as the user  interacts with  the machine. This  fluctuation  is due to the distortion from   metals and electronic chips installed  in these  machines forming  a peak in the magnetic field  readings (detected by the peak detector). Finally, as these  machines are usually  mounted to  walls, the passenger is forced  to change her direction  to walk away  as soon as  she finishes the service.  This instantaneous change  in  the user direction is reflected to a surge in the gyroscope readings when the user starts to resume walking (detected by the surge detector). This unique  patterns  are leveraged by  \sys{} to separate this type of machines from other  semantics (Figure \ref{other}). Now, we will  give the detail of how to discriminate the three classes of coin operated machines.
 \subsubsection{Drink Vending Machines} 
 Based on our observation, they  have   a unique loud sound emitted  when they are dispensing  drinks to the customer. This sound is  emanated  when the drink is pulled down from the machine storage into its  outlet. \\
 To verify that we can recognize the unique drink falling sound in the ambience, in our  preliminary  experiment, we recorded  an audio clip during the  usage of a drink vending machine (Figure 8). The time-domain audio signal contains coin insertion sound existing in all coin operated machines,  the drink  falling sound, and the background noise respectively (highlighted by the three consecutive blue boxes in Figure \ref{fig:ticketvsdrink}d). The Fast Fourier Transform (FFT) of the audio signal shows a clear peak at the 350Hz  frequency band in the drink falling audio clip (Figure \ref{fig:ticketvsdrink}c) while  no peaks  are evident at the 350Hz frequency band  neither in the coin insertion nor the background noise clips  due to the absence of this distinct acoustic signal (Figs. \ref{fig:ticketvsdrink}a, \ref{fig:ticketvsdrink}b). We observed that this frequency is consistent across all drink vending machines we experienced in the eight different stations in our dataset. \textit{During the system operation}, we use an empirical threshold  of  three standard deviations (i.e., 99.7\% confidence level of noise) to detect the  drink  falling  acoustic signal in the ambient sound  recorded during the usage of  coin operated machines. If the received audio signal strengths in the 350Hz  frequency band  exceeds the threshold, indicating that the sound level is significant at this frequency band (as signal strength is jumped significantly at this frequency band), the  system confirms the detection of a  drink vending machine. 
\subsubsection{Ticket Vending Machines} Similarly, they emit  a unique beep sound  many times during the  user interaction (e.g.,  pressing  a button,  indicating the  end of transaction, etc).  We envision that  this beep signal can be leveraged as a reliable discriminator as it is  absent in lockers where users  insert  coin, put luggage and   lock the drawer without any  distinctive sound. We  incorporate  the  same acoustic detection algorithm used to identify drink vending machines to separate ticket  vending machines from lockers. Figure \ref{fig:ticketvslocker}c shows  a raw audio recording  collected during buying a  ticket from a vending machine. We  crop two sections from the original audio signal comprising  the background noise and  the beep audio signal   respectively and convert these signals  into the frequency domain  by using FFT (Figs. \ref{fig:ticketvslocker}a, \ref{fig:ticketvslocker}b). We observed a clear peak  around the frequency of 3kHz  in the beep audio signal  whereas no peaks are observed at the frequency of 3kHz in the background noise  (Figs. \ref{fig:ticketvslocker}a, \ref{fig:ticketvslocker}b). When the ticket vending machine starts beeping, the signal strength in  the  3kHz  frequency band jumps significantly and therefore the ticket vending machine can be detected.
\subsubsection{Lockers}
To recognize lockers, we first attempted to identify the coin insertion sound  in the ambience  to avoid classifying all unrecognized activities as using lockers (i.e., catching all). However, we observed that while some coin sound signatures were visible, in many cases it was difficult to separate them from other frequency components (e.g., background noise). In addition, we find that the number of  traces, other than coin operated machines ones,  having  a stationary period accompanied with a significant magnetic distortion on the user's magnetometer  followed by a direction change (i.e., surge in gyroscope readings) are small. So, once vending machines samples are separated, the remaining  coin operated machines samples are classified  as lockers.
%%%%
  \begin{figure}[!t]
        \centering
        \begin{subfigure}[b]{0.23\textwidth}
                \includegraphics[width=\textwidth]{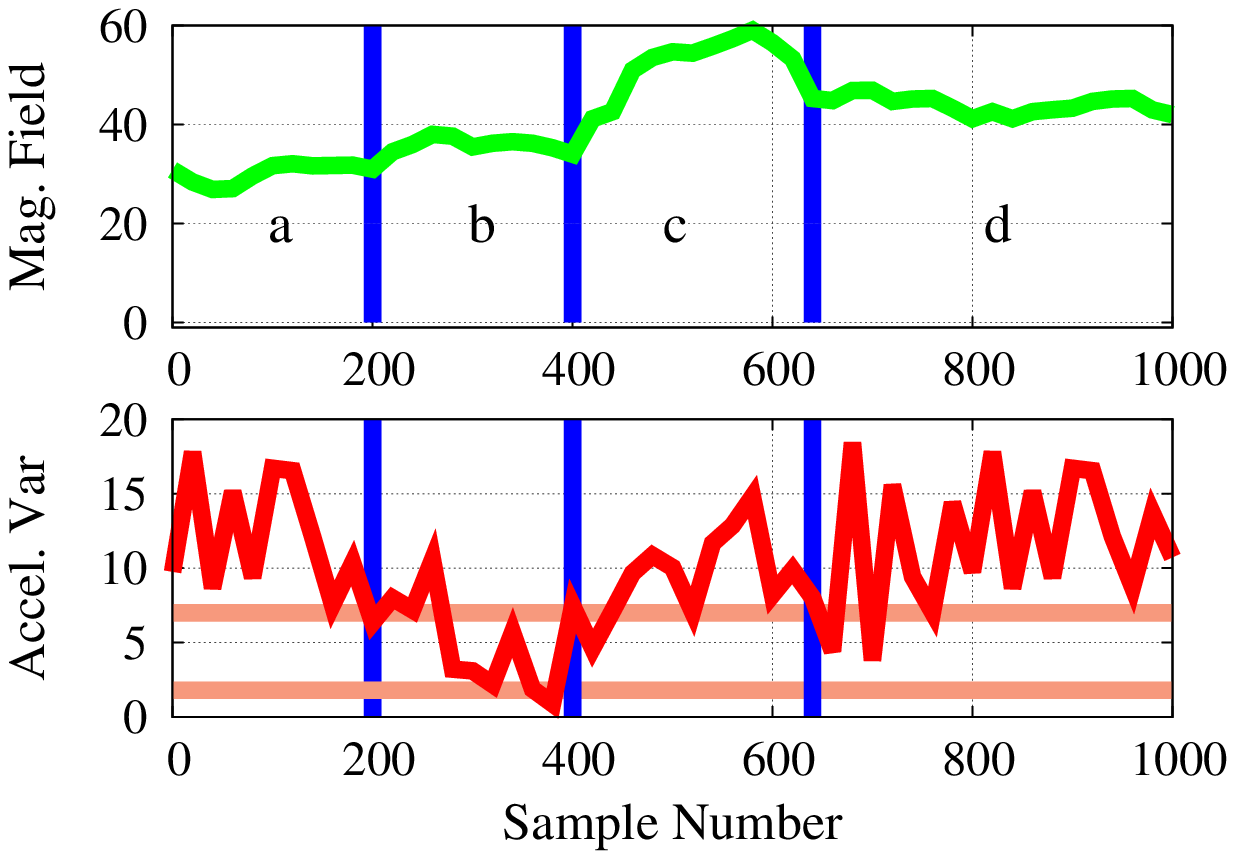}
           \caption{Ticket}
        \end{subfigure}%
        \begin{subfigure}[b]{0.23\textwidth}
                \includegraphics[width=\textwidth]{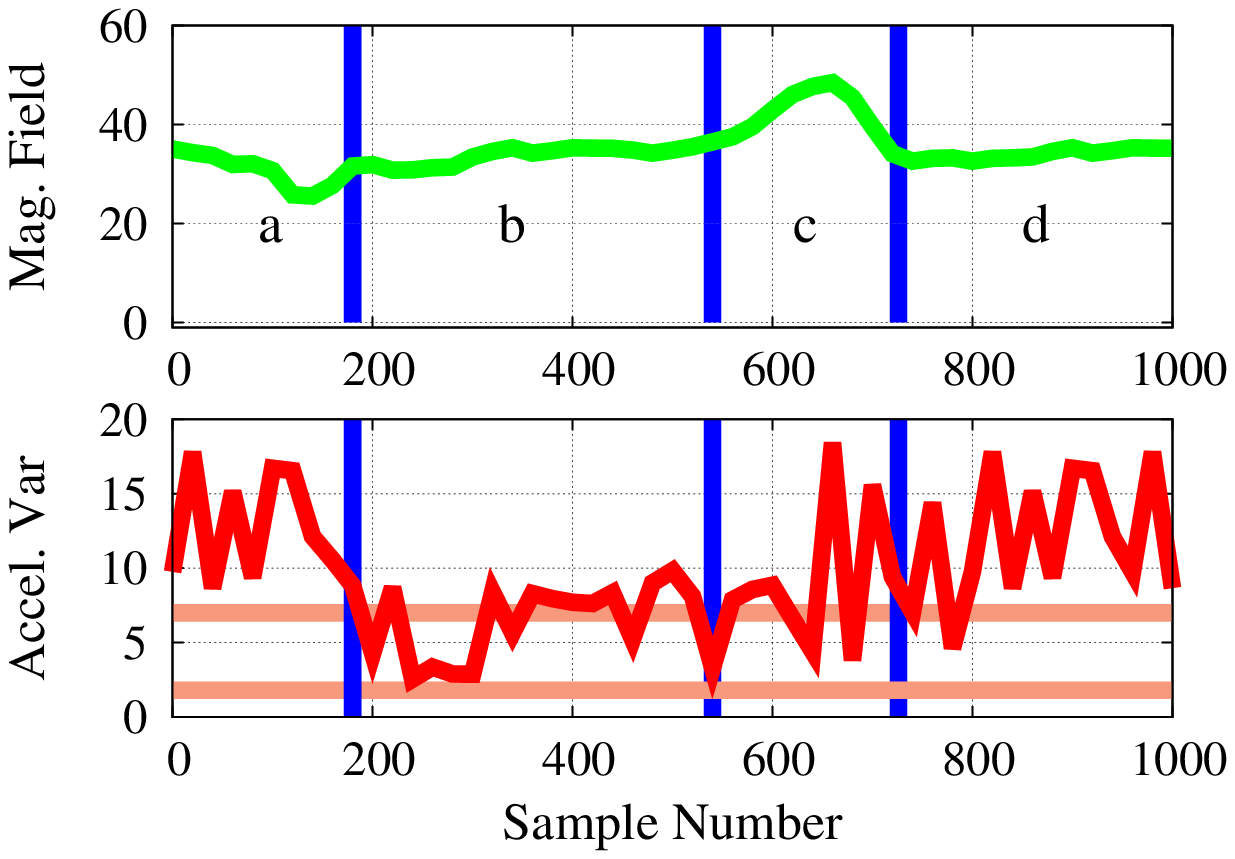}
              \caption{IC Card.}
        \end{subfigure}

        \caption{ The sensor pattern of crossing a gate by a ticket and an IC card. Both consist  of (a) normal walking, (b) deceleration near the gate, (c) acceleration accompanied by a peak on  ambient magnetic field, and (d) normal walking.}
\label{fig:ticketvsic}
\end{figure}
\subsection{Entrance Gates}
Railway passengers  have to pass through an automatic  fare collection gate  in their routes  to  the station's platform. To cross a gate, there are two ways:\\

\noindent \textbf{With a Ticket:}\\
Passengers mostly pass by a ticket vending machine to get a ticket. Thereafter,  as a passenger approaches the gate, a noticeable  slows down in her walking speed is observed till she  pauses  in front  of the  gate  to drop the ticket  into the  machine,  then  she  steps forward  to grab it from the machine, and finally she resumes normal walking (Figure \ref{fig:ticketvsic}a). This scenario translates to a unique motion  pattern consisting of the following sequence: normal walking, deceleration,  accelerating and normal walking. This unique motion pattern is detected by using the variance of acceleration (Figure \ref{fig:ticketvsic}a) where the two horizontal lines correspond  to the thresholds used to separate different motion patterns. Simultaneously, crossing the gate heavily distorts  the magnetic field by the gate ferromagnetic metals  forming a distinct peak  on the magnetometer reading (detected by a simple peak detector). \newpage

\noindent \textbf{With an IC Card:} \\ Nowadays, IC cards are commonly used for paying transit fees in many areas (e.g., Pasmo\footnote{http://www.pasmo.co.jp/en/} in Japan, Ventra\footnote{https://www.ventrachicago.com/} in Chicago). This  gate entrance method has  two differences from the ticket based one. First, passengers using IC cards do not have to pause as the  card reader can  recognize  the  card  while it is in close proximity  in users' hand or  wallet (acceleration variance  still above the stationarity threshold (Figure \ref{fig:ticketvsic}b)). Second, mostly  it is not preceded by the usage of a  ticket vending machine activity (i.e., neither sequential  nor dependent  activities).
\begin{figure}[!t]
        \centering
        \begin{subfigure}[b]{0.23\textwidth}
                \includegraphics[width=\textwidth]{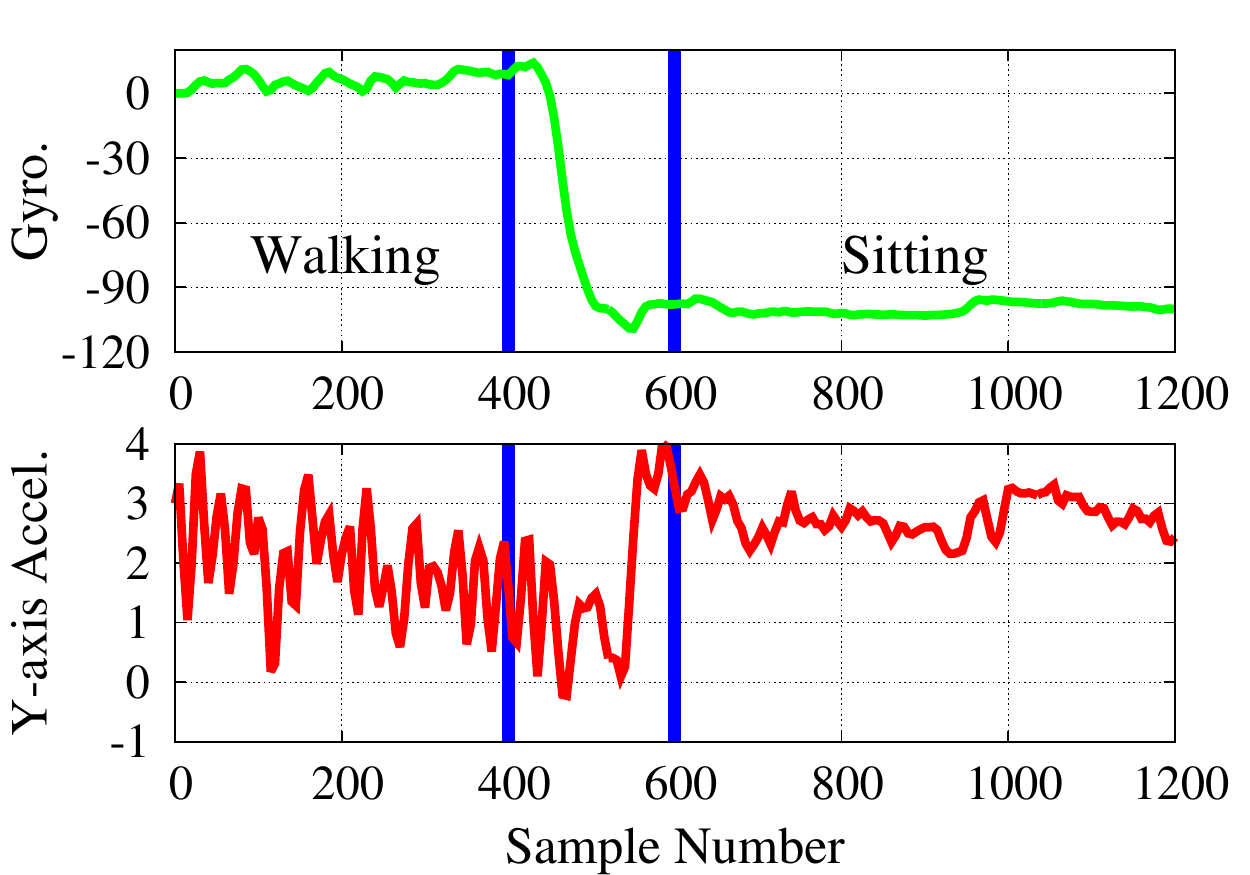}
           \caption{Sitting.}
        \end{subfigure}%
        \begin{subfigure}[b]{0.23\textwidth}
                \includegraphics[width=\textwidth]{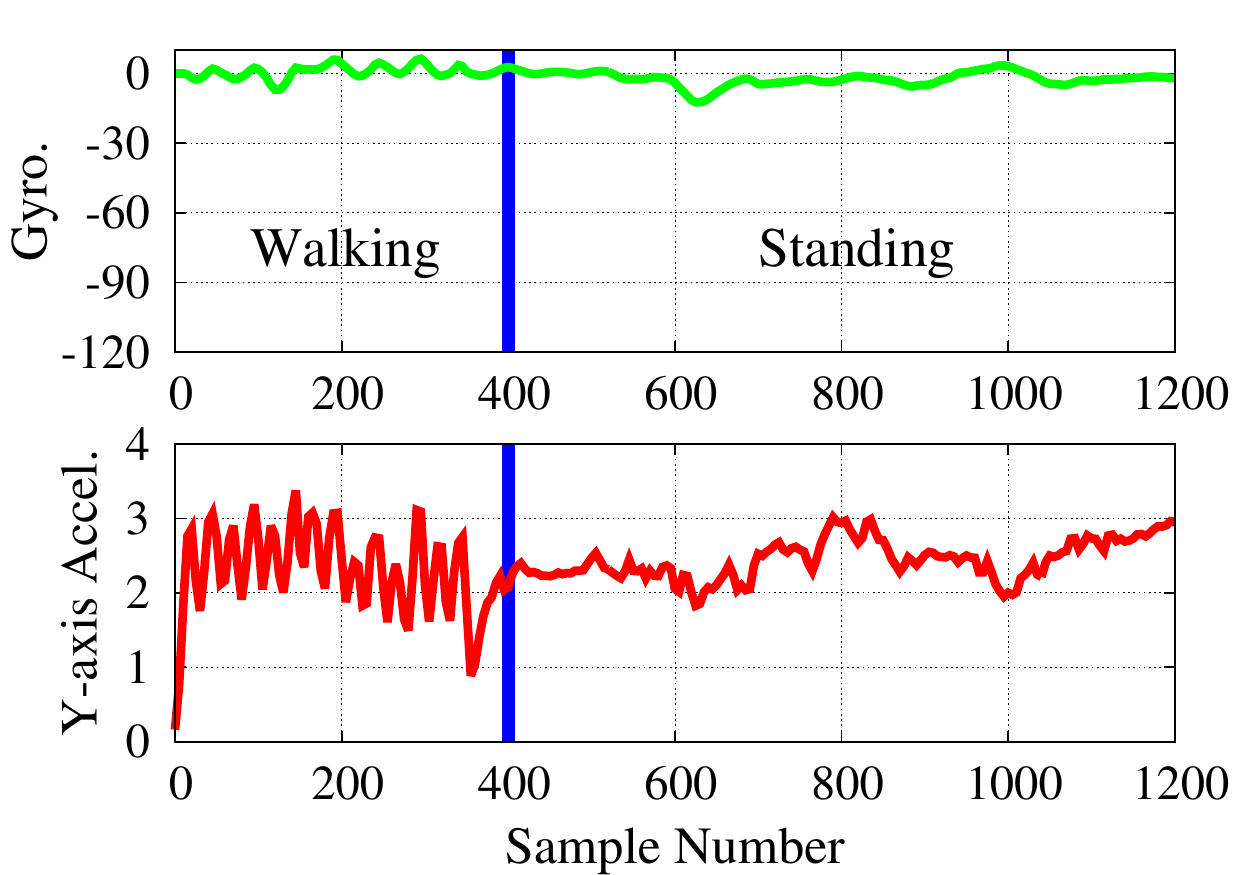}
              \caption{Standing.}

        \end{subfigure}

        \caption{ The effect of  activity transition from walking to sitting against the effect  of transition from walking to standing on the Y acceleration and gyroscope readings. }
\label{fig:sitting}
\end{figure}
\subsection{Waiting (Sitting) areas}
 Waiting areas are available in many stations platforms, especially those  where trains inter-arrival time is long.  Many people especially elderly people, pregnant  women, people with disabilities  and even normal passengers (e.g., in  the winter)  prefer to  wait for  trains  in this area.  We postulate that  if there are many activities transitions from walking to sitting taking place within  the premises  of a certain area, then this area will be  a waiting  area with a high probability.  To switch  from walking  to sitting, the passenger has to rotate first  to be  aligned with the seat  and then sit down. The instantaneous surge in gyroscope reading (from the direction change)  followed by a difference in relative magnitudes  of Y acceleration values (from forward and backward motions while sitting) are used to characterize the transition from walking to sitting (Figure \ref{fig:sitting}a). Conversely,  the  transition from walking to standing does not involve a noticeable difference on  the readings of Y acceleration (Figure \ref{fig:sitting}b). We leverage  the change in direction as an affirmative feature  to decrease the false positive rate in the sitting recognition given that the change in Y acceleration values may happen in  other conditions (e.g.,  normal  device bouncing while the user is  walking).
\subsection{Restrooms}
Public restrooms are available in almost all stations.  Normally, people  must have  stationary periods  while they are at restrooms (e.g., hand washing). Moreover,  due to the sensitive nature of public restrooms, their entry doors are faced by walls  so users  have to change their directions  after crossing entry doors. However, the  user stationarity and  direction change  patterns are not sufficient to efficiently separate the user being in a restroom from  other contexts. To  accurately identify restrooms, we incorporate the algorithm in \cite{fan2014public}  which detects  restrooms by actively probing the acoustics of environment with the built-in speaker and  microphone on the mobile phone. A probing sound is emitted by the phone and the impulse response (IR)  is analyzed to detect the type of  space (restroom or not). The acoustic  characteristics of an environment depend on its dimension and its ability to absorb sound. Since, public restrooms have similar affordances (e.g., water resistance floors and wall, toilets and sinks), they have a unique absorption coefficient of sound and thus they can be detected easily. Since the sweep volume level does not affect the accuracy of the model significantly \cite{fan2014public}, \sys{} leverages sweeps with lower volumes to avoid being invasive. Additionally, the model prediction performance is  robust against
the restroom occupancy and sounds generated by the occupants
(e.g., flushing or hand-washing) \cite{fan2014public}.
\subsection{Platform and Waiting Lines}
\begin{figure}[!t]
\centering
\begin{minipage}[t]{0.48\linewidth}
\includegraphics[width=1\textwidth,height=3cm,keepaspectratio]{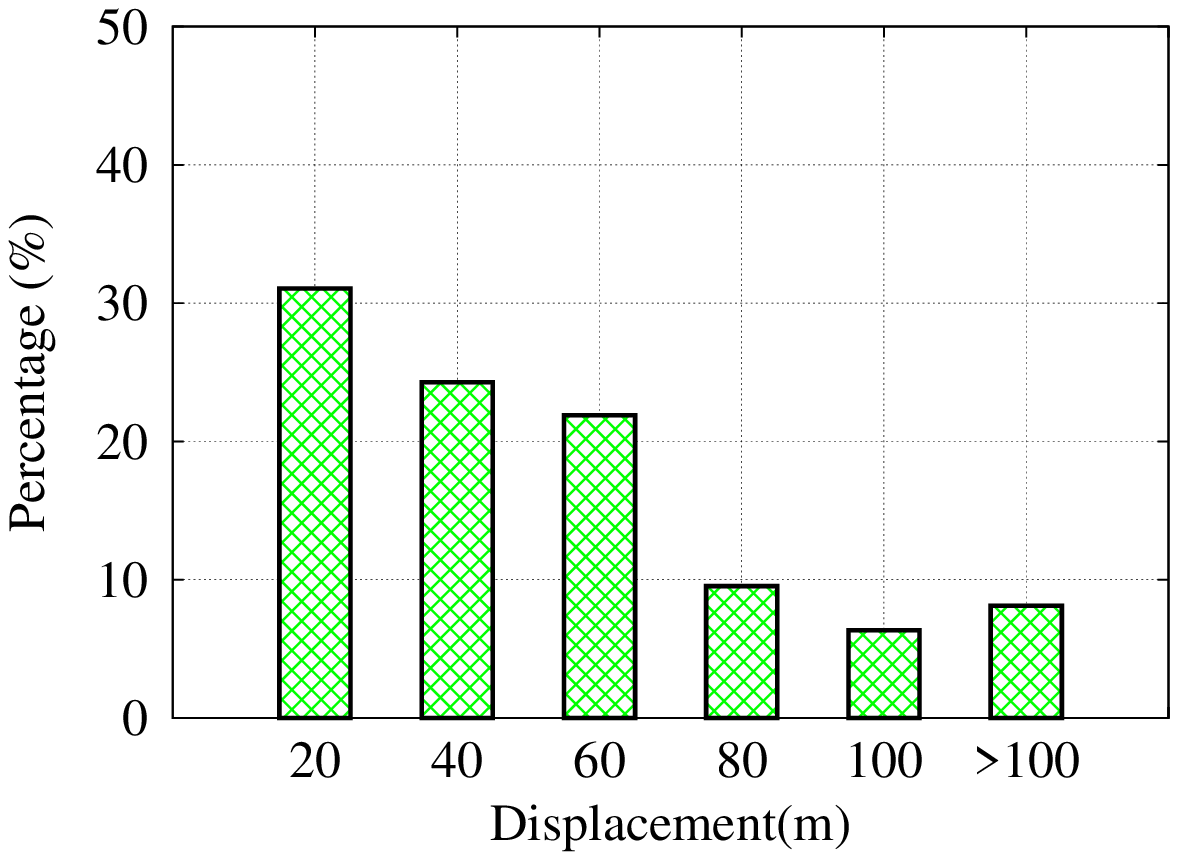}
\caption{Passengers displacement distribution over the platform.}
\label{passenger}
\end{minipage}
\hfill
\begin{minipage}[t]{0.48\linewidth}
                  \includegraphics[width=1\textwidth,height=3cm,keepaspectratio]{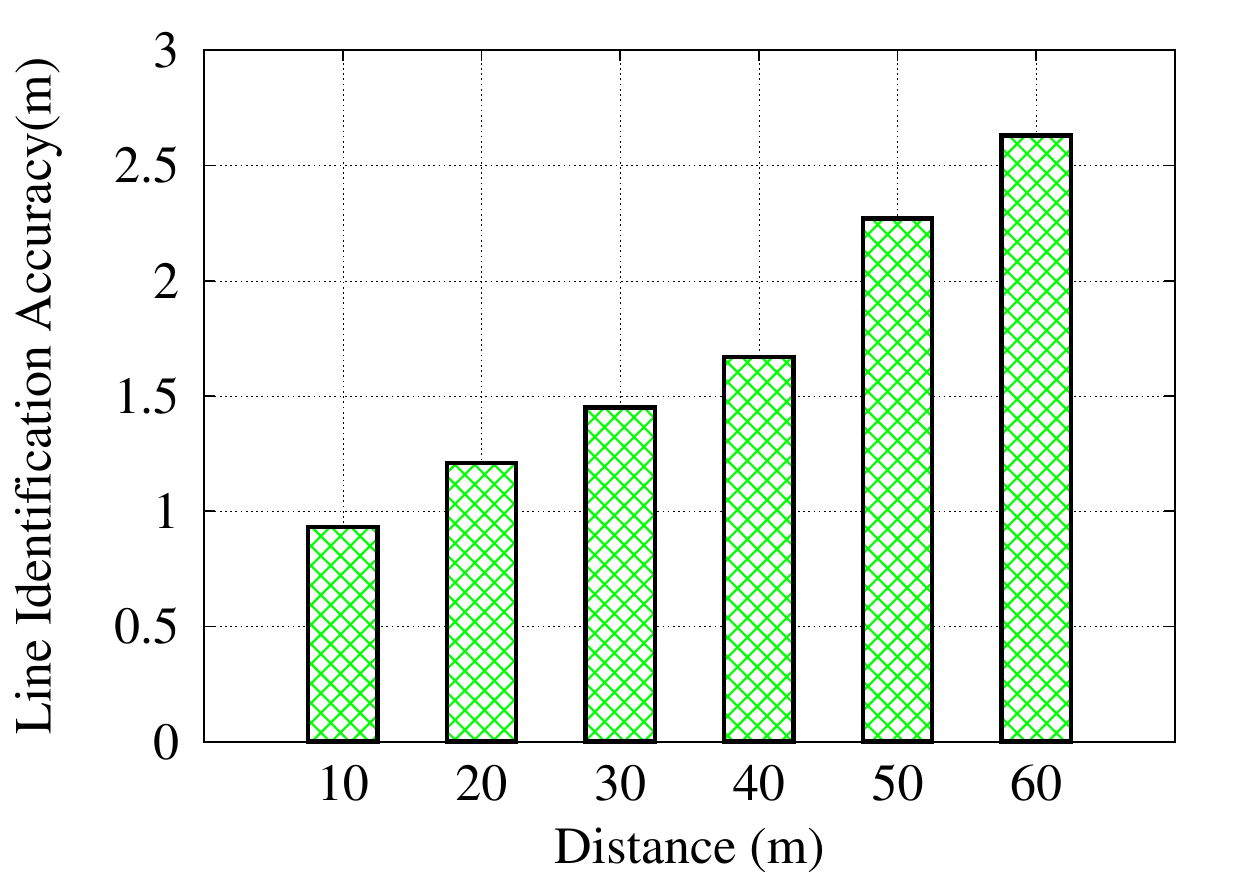}
                  \caption{Lines location accuracy versus their distance from the platform access}
                  \label{line}
\end{minipage}
\end{figure}

To further enrich the semantics of \sys{}, we also identify the platform area and the location of waiting lines for train cars. The platform
 is the area where passengers board the train. Therefore,  their transportation mode changes from standing (waiting for train)  to walking (into the train car)  to be in a  motorized transport (when the train moves) during a short time period. Transportation mode detection has been thoroughly studied in literature, e.g. \cite{hemminki2013accelerometer,abdelaziz2015diversity,krumm2004locadio,reddy2010using,sohn2006mobility}. 
 We follow the approach proposed in \cite{hemminki2013accelerometer} that provides high accuracy of transportation mode detection (walking, stationary or in a motorized transport) based on the energy-efficient accelerometer sensor. The short temporal consequence of  standing, walking shortly and being in a train activities is leveraged to detect the platform track and its position is estimated from passengers' positions during train boarding.\\
 Once the platform area is identified, the waiting line positions can be estimated  from  the passengers' positions reported while they are waiting for  the trains (i.e., switching from walking to standing activity on the platform (Figure~\ref{fig:sitting}b)). We videotaped  a sample of 3000 passengers' traces starting  for their access to the platform (escalator, elevators and stairs) until they join a waiting line. Figure \ref{passenger} shows that  about 76\% of  passengers walk  a short distance (less than 60m) on the platform  where they tend to join the nearest uncrowded waiting line. This is partially due to the fact that platforms are typically designed  to have multi-accesses to disperse the passengers load. In addition, as passengers  use  elevation change semantics to access the platform, their dead-reckoned derailed position will be curbed and calibrated to the access locations of these semantics on the platform when this semantic surface on the user's trace.
 This verifies that \sys{} can localize the user accurately on the platform. Thus, waiting lines positions can be estimated by the clustering approach from  passengers' locations when  they are  waiting in a line for the train. Once lines are detected, every neighbored collection of $n$ lines (where $n$ is the number of doors per car, which is a known constant) are representing a queuing  area for a train car. 
\section{ Evaluation}
\label{sec:eval}
\sys{} is  evaluated through a  deployment  at  eight different stations of different sizes  in two different  cities (Osaka and Kobe)  in Japan.
Table \ref{dataset} shows the detailed description of the collected dataset. The  stations are managed by different companies; having different  buildings designs  and sizes; and  semantics placement. This emphasizes the scalable nature of \sys{}. The average length and width of platforms are 160m and 12m on average respectively. Train cars are of 20m length with 3 doors  and  the average inter-door distance is 5.2m.
 \begin{table*}[!ht]
\centering
%\tiny
\caption{The description of collected data.}
\label{dataset}
\resizebox{\linewidth}{!}{
\begin{tabular}{||c|| c|c| c| c|c|c|c|c|c|c|c|c|c|}
  \hline
            \textbf{ Station}   &\textbf{Osaka}&\textbf{Umeda}&\textbf{Senrichuo}&\textbf{Juso}&\textbf{Higashi-Umeda}&\textbf{Nishinomaya} &\textbf{Kobe-Sanyomia}&\textbf{Shin-Kobe}\\\hline
\textbf{No. of Traces}&108&78&71&69&61&59&53&48 \\\hline
\textbf{No. of Users}&10&9&7&6&6&6&5&4\\\hline
\textbf{Type and City}&JR- Osaka&Subway- Osaka&Subway- Osaka&Hankyu- Osaka&Hankyu- Osaka&Hankyu- Osaka&Subway- Kobe&Subway- Kobe\\\hline
\end{tabular}
}
\end{table*}

\subsection{Data Collection Methodology}
A group of 16 volunteers,  of different ages (11 in their  20's and 5 in their 30's) and gender (12 males and 4 females),  collected  the necessary data for evaluation. The collected data consists of two datasets-\textbf{\textit{scenario based and free}}- differing in how the data is collected. In the scenario-based dataset, 10 participants were assigned  specific trajectories  starting from the station entrance to different platforms. The  trajectories were selected carefully to cover all possible routes that  were exhibited  by daily passengers while covering all available semantics  at the same time. While  heading to a specific platform, some participants purchased  tickets to cross entrance gates;  others crossed gates directly using  different types of IC cards; some participants  used  drink vending machines while others used  lockers; and all participants passed through elevation change semantics to access the platform. On the other hand, the free dataset is collected by six individuals from their everyday train commutation in different stations. \\
We have deployed  two Android  applications: The first application  is a data collection tool  that runs in the background to sample all inertial sensors, the  barometer at 50Hz as well as recording audio. The second application is designed  for ground truth collection and runs in the foreground  to allow participants to  manually  tag their  activities. The data collection was conducted in  different times, different days and  using different Android phones including Samsung Galaxy S5 and LG Nexus 5. Participants carry smartphones  in different placements (in hand  or in the trouser pocket). When the participants carry the logger phones in the trouser pocket, they carry another phone in their hands to annotate their activities.  This  captures the time-variant nature of semantics signatures and stations congestion level; generalization of the system over different stations;  as well as the heterogeneity of users and devices.

\subsection {Performance Results}

In this section, we evaluate  the  semantics identification accuracy, semantics  location estimation accuracy,   power consumption,   impact of different phone placement and finally quantify the generalization performance of \sys{}.

\subsubsection{Semantics Detection Accuracy}
We evaluate  the  semantics detection accuracy  based on  the scenario-based datasets. The detection accuracy is measured by  false positive \footnote{Samples of other semantics  classified as the current semantic.}  and false negative \footnote{Samples of the current semantic that are not detected.} rates.

Table~\ref{levelchangeconf} shows  the confusion matrix for  detecting various elevation change semantics. It shows that  some elevation change semantics are easy to detect  due to their unique patterns. This leads to zero false positive and  false negative rates for the elevators (the coarse-grained category), half-landing stairs, and escalator  (when users are standing) cases. However,  climbing straight  stairs   sometimes are misclassified as  climbing escalators when the stairs are very close to escalators so they have a similar magnetic distortion signature. Moreover,  separating the two types of elevators  generates some misclassifications which is due to the different  elevator usage patterns (e.g., some users of single-door elevators do not  turn around completely in one step). Nevertheless,  as standing in or climbing  up  escalators activities are translated to the same semantic (escalator),  \sys{} can still achieve a high detection accuracy for elevation change semantics with  3.3\% false positive and 4.1\% false negative rates on overage.

\begin{table*}[!ht]
\centering

\caption{Confusion Matrix for classifying  different  \textbf{elevation change} semantics.}
\label{levelchangeconf}
\resizebox{\linewidth}{!}{
\begin{tabular}{||c||c|c|c|c|c|c|c||c|c||c||}
  \hline
   &Elevator (Single)&Elevator (Double)&Stairs (Straight)&Stairs (Half-land.)&Escalator (Stand.)&Escalator (Climb.)&\cellcolor{Gray}Escalator (Over.)&FP&FN&$\sum$\\\hline
    Elevator (Single)& \cellcolor{Yellow}\textbf{100}& 5&0&0&0&0&\cellcolor{Gray}0&   6.7\%&4.8\%&105\\\hline
  Elevator (Double)&7&\cellcolor{Yellow}\textbf{79} &0&0&0&0&\cellcolor{Gray}0&   5.8\%&8.1\%&86\\\hline
  Stairs (Straight)&0&0 &\cellcolor{Yellow}\textbf{111} &0&0&9&\cellcolor{Gray}9&    0\%&7.5\%&120\\\hline
  Stairs (Half-land.)& 0&0 &0&\cellcolor{Yellow}\textbf{51} &0&0&\cellcolor{Gray}0&    0\%&0\%&51\\\hline
    Escalator (Stand.)& 0&0 &0&0&\cellcolor{Yellow}\textbf{114}&0&\cellcolor{Gray}   -&-&-&114\\\hline
  Escalator (Climb.)&0 &0&0&0&0&\cellcolor{Yellow}\textbf{107} &\cellcolor{Gray}     -&-&-&107\\\hline
   \hline \cellcolor{Gray} Escalator (Over.)&\cellcolor{Gray} 0&\cellcolor{Gray}0&\cellcolor{Gray}0&\cellcolor{Gray}0&\cellcolor{Gray}-&\cellcolor{Gray}-&\cellcolor{Yellow}\textbf{221} &    4.1\%&0\%&221\\\hline
  \hline Total&   \multicolumn{7}{c||}{} &     3.3\%&4.1\%&583\\\hline
\end{tabular}
}
\end{table*}

Table~\ref{coin} shows  the confusion matrix  representing  the detection accuracy of station specific semantics  that  force users   to be stationary while using them (i.e., semantics in the right branch of the decision tree in Figure  \ref{other}). It shows that  coin operated machines (the coarse-grained category) can be  detected  with  near 100\% accuracy using their unique inertial sensors pattern. To classify coin operated machines to their fine-grained categories (drink vending machines, ticket vending machines and lockers), the acoustic based detection scheme can achieve a good  accuracy with  10.4\% and 6.9\%   for false positive and  false negative rates respectively on average. This is due to  the  responsive sound of vending machines that is universally used in many  stations in Japan\footnote{Vending machines around Japan are similar in their user interface and hardware to facilitate the  Human-Machine interaction as well as  to be able to recognize  the same IC card types used to pay transit fees and drinks cost across the country.}, making it  an identifiable signature for classification.  On the other hand, standing and waiting for  train activities are sometimes interchangeably misclassified. For example, when a user is standing on the platform (to answer a phone call), this may be  interpreted as waiting for train and oppositely the user may be  waiting for train but it is misclassified as standing when the system failed to recognize that the user is on the platform. \sys{} lessens this effect by using the transportation mode detection algorithm to detect when  the train has moved, which comes after the waiting for train inactivity, as opposed to any other type of inactivity.
\begin{table*}[!ht]
\centering
\caption{Confusion Matrix for  classifying different stations specific semantics discovered  from \textbf{stationary}  traces.}
\label{coin}
\resizebox{\linewidth}{!}{
\begin{tabular}{||c||c|c|c|c|c|c|c||c|c||c||c||}
  \hline
   &Drink Vending Machine &Ticket Vending Machine&Locker&Restroom& Waiting (Sitting) Area&Standing&Cars' Waiting Lines&FP&FN&$\sum$\\\hline
  Drink Vending Machine & \cellcolor{Yellow}\textbf{123}  &1&6&0&0&0&1&   7.6\%&6.1\%&131\\\hline
  Ticket Vending  Machine&9&\cellcolor{Yellow}\textbf{260} &12&0&0&0& 0&     3.2\%&7.5\%&281\\\hline
 Locker&1&8&\cellcolor{Yellow}\textbf{119 } &0&0&0&0&       20.3\%&7\%&128\\\hline
Restroom&0&0&4&\cellcolor{Yellow}\textbf{87}&2&4&  0  &   7.2\%&10.3\%&97\\\hline
Waiting (Sitting) Area&0&0&0&4&\cellcolor{Yellow}\textbf{88} &6&  3 & 11.9\%&12.9\%&101\\\hline
Standing&0&0&3&3&7&\cellcolor{Yellow}\textbf{140} &  5   &  11.4\%&11.4\%&158\\\hline
Cars' Waiting lines (waiting for a train)&0&0&1&0&3&8&\cellcolor{Yellow}\textbf{125}    &   6.6\%&8.8\%&137\\\hline
  \hline Total& \multicolumn{7}{c||}{} &            9.7\%&9.1\%&1033\\\hline

\end{tabular}
  }
\end{table*}

 \begin{table*}[!ht]
\centering
\caption{Confusion Matrix for  classifying  different stations specific semantics   discovered from  \textbf{non stationary} traces.}
\label{gate}
\resizebox{\linewidth}{!}{
\begin{tabular}{||c|| c|c| c| c|c||c|c||c||}
  \hline
   &Walking&Platform Track (Boarding)&Entrance Gate (Ticket)&Entrance gate (IC Card)& Entrance gate (Overall)&FP&FN&$\sum$\\\hline
  Walking&\cellcolor{Yellow}\textbf{170}&7 &5&9&\cellcolor{Gray}14&10.5\%&11\%&191\\\hline
  Platform Track (Boarding)&3&\cellcolor{Yellow}\textbf{135} &2&11&\cellcolor{Gray}13&15.2\%&10.6\%&151\\\hline
  Entrance Gate (Ticket)&9&2 &\cellcolor{Yellow}\textbf{264} &6&\cellcolor{Gray}-&-&-&281\\\hline
   Entrance gate (IC Card)&8&14&9&\cellcolor{Yellow}\textbf{234} &\cellcolor{Gray}-&-&-&265\\\hline
 \hline \cellcolor{Gray}  Entrance gate (Overall)&\cellcolor{Gray}17&\cellcolor{Gray}16&\cellcolor{Gray}-&\cellcolor{Gray}-&\cellcolor{Yellow}\textbf{513} &4.9\%&6.0\%&546\\\hline
  \hline
  \hline Total&   \multicolumn{5}{c||}{} &10.2\%&9.2\%&888\\\hline
\end{tabular}
}
\end{table*}
 \begin{table*}[!ht]
\centering
\caption{The semantics classification accuracy in a \textbf{one-station-out} cross validation.}
\label{cross}
\resizebox{\linewidth}{!}{
\begin{tabular}{|l|| l|l| l| l|l|l|l|l|l|l|l|l|l||l||}
  \hline
                &\textbf{Elevator(S.)}&\textbf{Elevator (D.)}&\textbf{Stairs (Str.)}&\textbf{Stairs (Half.)}&\textbf{Escalator}&\textbf{Drink Vd. Mch.}&\textbf{Ticket Vd. Mch.}&\textbf{Lock.}&\textbf{Restr.}&\textbf{Sit. Area}&\textbf{Wait. Lines}&\textbf{Plat. Tr.}& \textbf{Entr. gate}&$\sum$\\\hline   \hline
\textbf{FP}&8.6\%&4.6\%&0\%&0\%  &5.4\%            &9.1\%  &3.9\%&21\%    &8.2\%&10.8\%  &8\%           &9.2\%       &5.3\% &7.2\%    \\\hline
  \textbf{FN}&3.8\%&10.4\%  &10\%&0\%  &0\% &   7.6\%  &8.2\%&  8.6\%  &9.3\%&   15.8\% &10.2\%     &11.9\%    &8.6\%  &8.0\%\\\hline
 \end{tabular}
}
\end{table*}
The confusion matrix of  classifying semantics from non stationary traces (i.e., semantics in the left branch of the decision tree in Figure  \ref{other}) is shown in Table~\ref{gate}. The table shows that \sys{} can reliably detect crossing  entrance gates using  a ticket.  The detection of  crossing of entrance gates  by  IC cards is a bit challenging   as some passengers either  do not  slow down  their walking sufficiently  or walk very  slowly  making their signature similar to ticket-based methods. Even worse, sometimes the card reader does not recognize the IC card  and  the passenger has to rollback and cross the gate again.
However, since  all  gates have IC card readers and  ticket slots integrated into the same machine,  the two entrance methods (IC card and ticket)  are  aggregated into one semantic (entrance gate)  that can be identified with 5.9\% and 6\% false  positive and false negative rates, respectively. Moreover, the detection of train boarding is sometimes misclassified as  crossing a gate by  an IC card. The main reason is that train motors and electrical inverters emit large magnetic noise during the acceleration periods of the train which sometimes coincides with the  pattern arising when the user crosses a gate using an IC card.
To reduce this, \sys{} leverages the sound emitted when the IC card touches the reader. Nevertheless, there is a trade-off between energy consumption and the semantic detection accuracy in this case.\\
Figure \ref{line} reports the accuracy of detecting the waiting line locations, as computed by \sys{}. Aligned with our intuition, the accuracy of waiting line locations detection  relies on their placement  with respect to the platform accesses. Due to the limited area of platform (average dimension is 160m$\times$12m), the short movement of a user on the platform (Figure \ref{passenger}), and the average inter-distance between waiting-line (5.2m);  \sys{} can estimate the line positions accurately, especially those near to the platform accesses where most user trails are short and thus the location accuracy is high  \cite{abdelnassersemanticslam,wang2012no}.\\
 Finally,   \sys{} can consistently  detect  the fine-grained classes of semantics accurately with 7.7\% false positive rate and 7.5\% false negative rates on average.
\subsubsection{Discovered Semantics Location Accuracy}
In this subsection, we study how much data is enough for \sys{} to  estimate  semantics  locations accurately as in crowdsensing-based systems the accumulation of more  samples  will enhance the  system performance.  Figure \ref{fig:location} quantifies the effect of the number of crowd-sensed samples on the accuracy of semantics (apart from waiting lines). The figure  shows that even if some semantics have some outliers, the system can achieve a good accuracy in estimating  their locations. This stems from the fact that independent correct  samples  of the same semantic are  in adjacent  locations  and tend to cluster while erroneous samples are widely scattered in the spatial space and do not form a cluster. In addition, \sys{} works offline so as a user encounters a semantic, \sys{} learns her errors, and therefore can track back and partly correct her past trail thus  the effect of the cold start problem is mitigated.  Finally,  as stations are  rich with semantics,  the localization error grows and sharply drops at semantics curbing the localization error and in its turn enhances the semantic location accuracy. Even though the instantaneous PDR error still  has an effect on the semantics locations estimation,  especially when the number of samples of semantics are small, it is evident from the figure that this error will  drop quickly as the number of crowd-sensed samples increases.  \sys{} can consistently achieve the accuracy of  2.5m using as few as  40 samples  for each discovered semantic type.  Thus, \sys{}  converges reasonably
quickly. However, we note that it needs to be periodically run to handle dynamic environment changes.
\subsubsection{Energy Consumption}
\begin{figure}[!t]
\noindent\begin{minipage}[t]{0.48\linewidth}
  \includegraphics[width=1\textwidth,height=2.6cm]{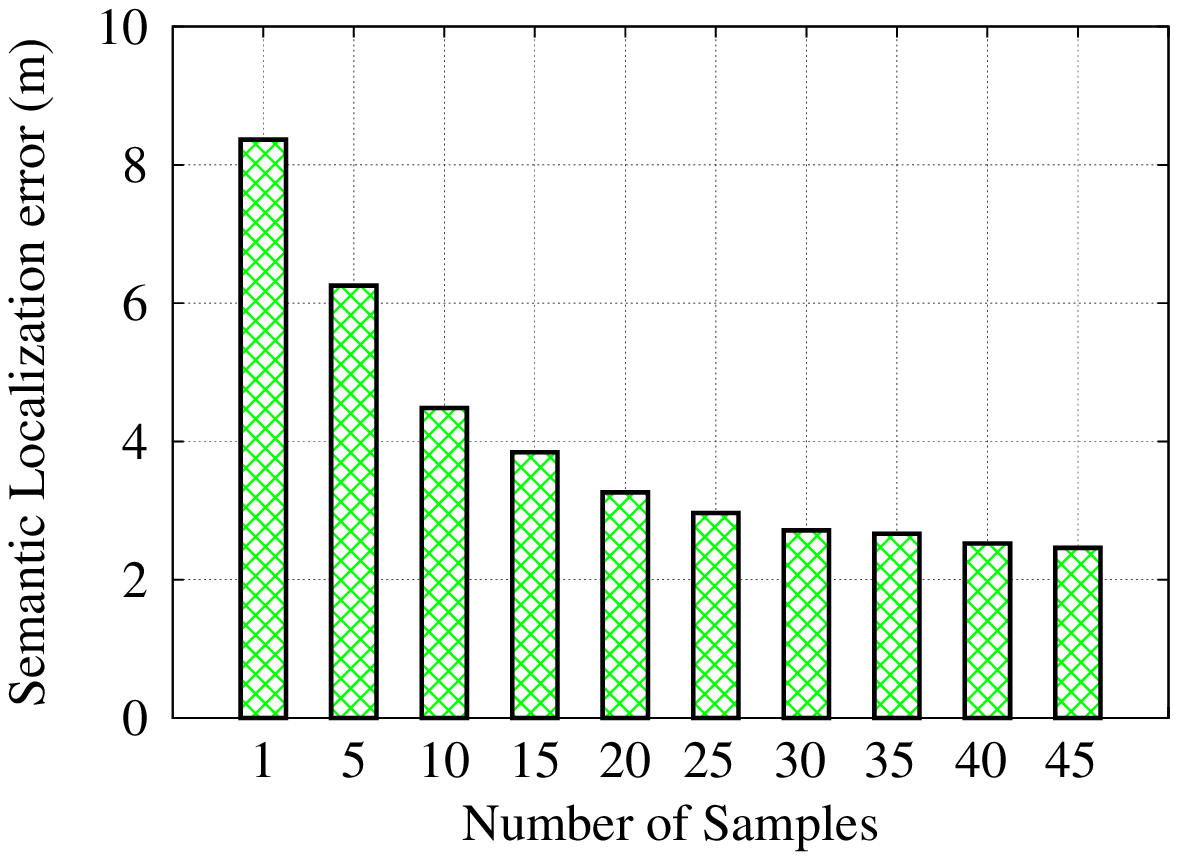}
  \captionof{figure}{Effect of the number of samples on the accuracy of semantics location estimation.}\label{fig:location}
\end{minipage}
\hfill
\begin{minipage}[t]{0.48\linewidth}
\includegraphics[width=1\textwidth,height=2.6cm]{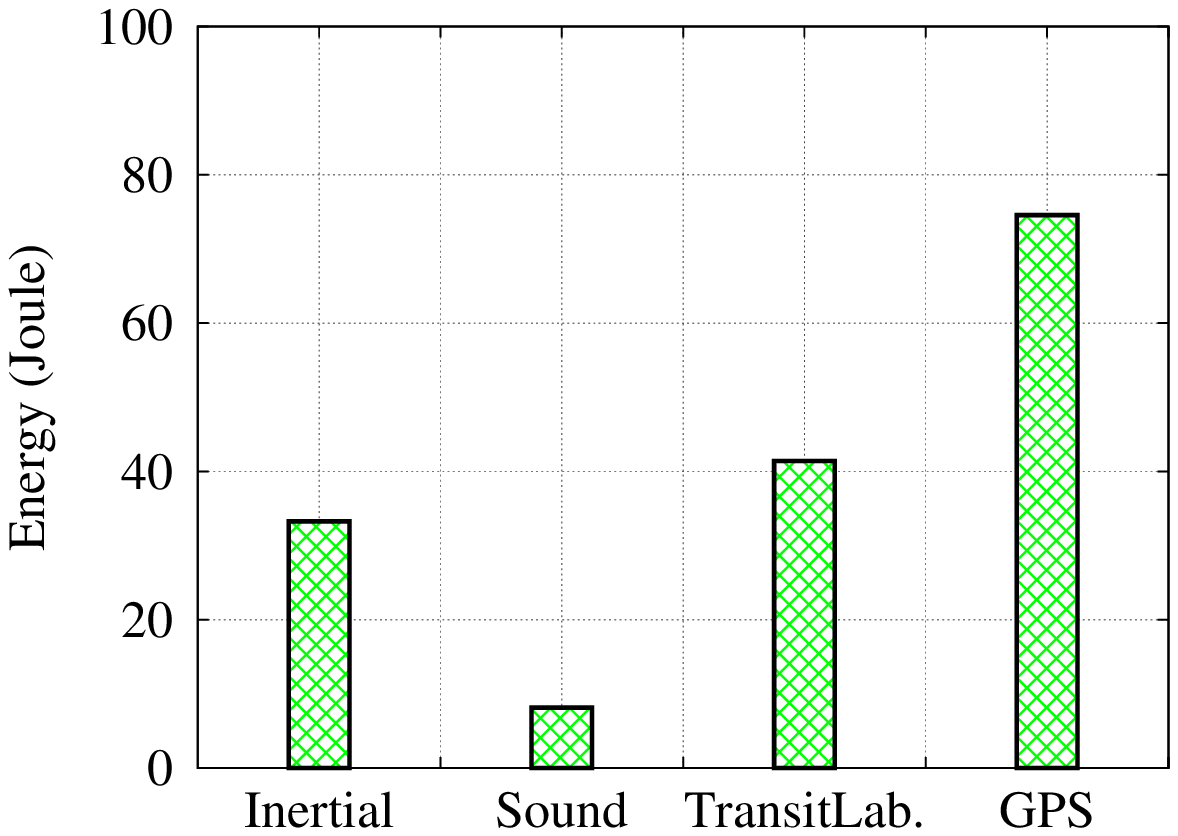}
\captionof{figure}{Energy footprint of \sys{}.}
\label{fig:ener}
\end{minipage}
\end{figure}

  Figure \ref{fig:ener}  shows the energy consumption of  \sys{} averaged over typical traces from entering the station to boarding trains. For this, we run an application that samples  the GPS every second to show the contrast  in power consumption (\emph{GPS is neither available in all locations in stations nor  able to detect  semantics but it is used as a baseline system}). The energy is calculated using the PowerTutor profiler \cite{gordon2013powertutor}  and the Android APIs using the HTC Nexus One cell phone.  \sys{}  leverages  the  inertial  sensors for  passengers' activity recognition and position estimation. Since inertial sensors are indeed used during the normal phone operation, to detect the phone orientation change  or estimate the user location for any indoor LBS, \sys{} practically consumes little \textbf{extra} sensing power in addition to the standard phone operation.
   In addition, the sound  sensor; which  has a higher energy footprint; is activated only  shortly during the  time of activities (average activity duration is  short- 31seconds- excluding the restroom). This also  avoids the impact of false positive triggers of the sound sensor (e.g., sensor being activated while the user is standing doing a phone call). Finally, as soon as the user waits for the train on the platform, we  suspend the activation of the  sound sensor.  All lead to a low energy consumption of \sys{} that is significantly (44\%) less than the GPS consumption.
\subsubsection{Impact of the Phone Placement}
\begin{figure}[!t]
\noindent\begin{minipage}[t]{0.48\linewidth}
  \includegraphics[width=1\textwidth,height=2.6cm]{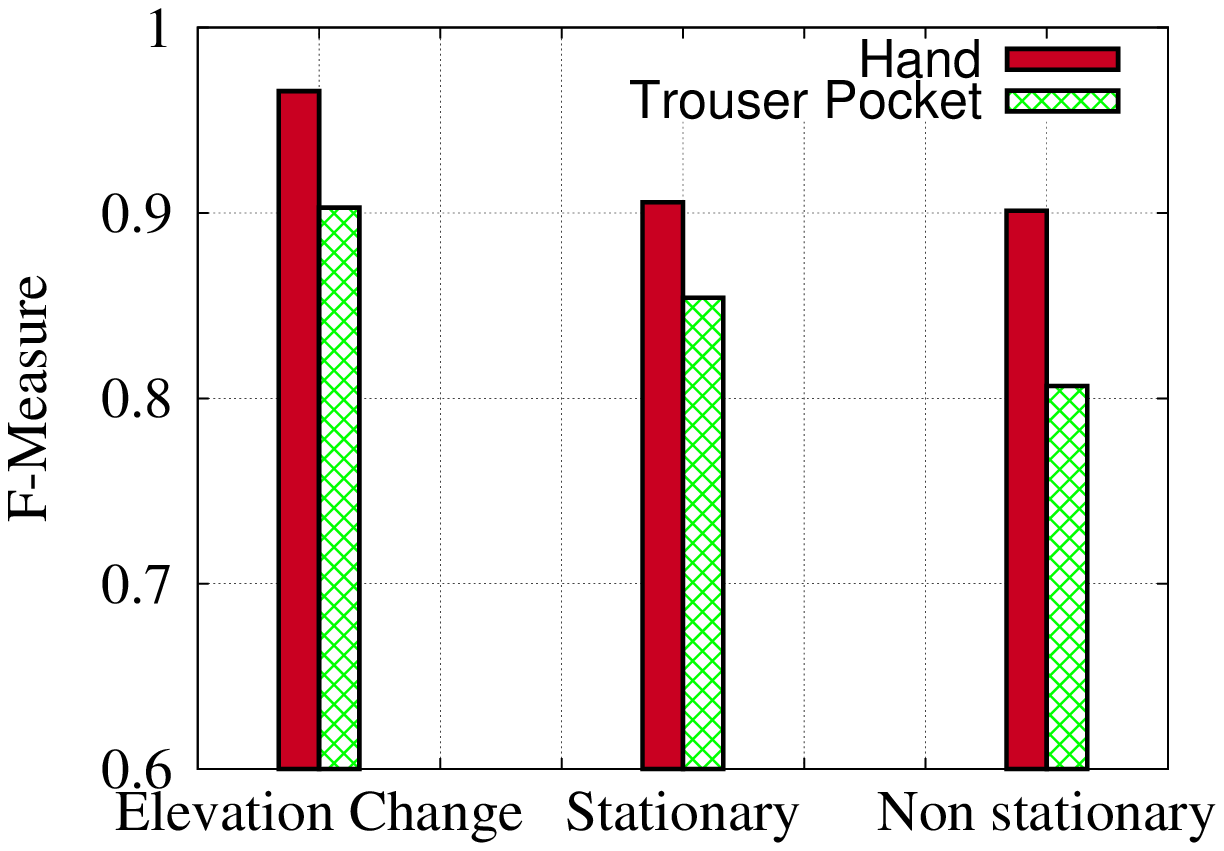}
 \caption{The Impact of phone placement on the semantics detection accuracy.}
\label{placement}
\end{minipage}
\hfill
\begin{minipage}[t]{0.48\linewidth}
\includegraphics[width=1\textwidth,height=2.6cm]{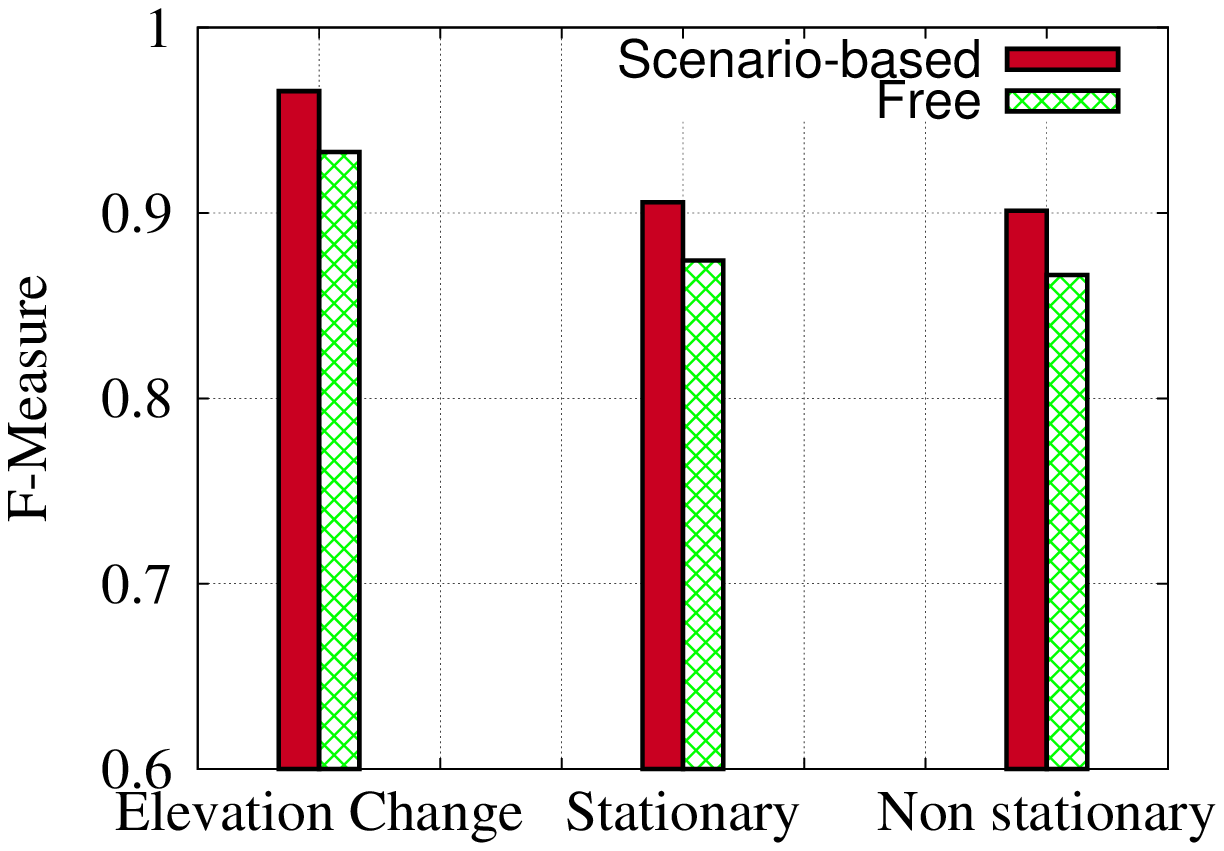}
\caption{The generalization experiment of \sys{} semantic detection.}
\label{fig:gen}
\end{minipage}
\end{figure}

To demonstrate that our approach is robust to different device placements, we carried out experiments on different phone placements (in a user hand or in a trouser pocket). Figure~\ref{placement} shows the semantics detection accuracy as measured by the F-measure \footnote{F-measure is the harmonic mean of precision and recall and represents a single number for comparison.} in both  phone placements. The results suggest that the semantic accuracy is not significantly dependent on device poses, with \sys{} achieving a high accuracy of about  80\% at least (in the case of non-stationary traces which is the most tricky case due to the effect of the legs movement on the phone in the pocket reading). \textit{This robustness mainly comes from the transformation of sensor readings to  the real world coordinates and the  extraction of placement-independent features from phone sensors}. For example, the magnetometer readings features are not largely attenuated by human bodies (i.e., in pocket). While the acceleration can be affected by the motion noise to some extent, its impact is effectively mitigated by smoothing the raw acceleration and  averaging the acceleration variance over a long time window. Finally, the collected audio  data is not much derailed by the phone placement in pocket as  passengers are forced to be  close to semantics at the usage time. Note that \sys{} uses different thresholds values to identify passenger's activities depending  on the phone placement (pocket or hand). Nonetheless, the light proximity sensor can be used to detect the phone position and, accordingly, decide which thresholds value to be used. Other phone positions can be handled in a similar manner, which is a subject of future work for space constraints.

\subsubsection{Generalization of TransitLabel}
  We based our  semantics identification on their typical usage pattern/signature by the majority of train commuters. However, some users may have different  usage patterns (regular versus occasional travelers)  and some stations  have different building structures and different machines  hardware  (e.g., vending machines) that may lead to  some semantics misclassifications. Thus,  to demonstrate  that \sys{}  could generalize over various users  and  stations, we consider an experiment  where a group of daily train commuters are asked to collect data  \textbf{\emph{freely (without prescribed scenarios)}} in different stations in two different cities managed  by different train companies in Japan.  Figure~\ref{fig:gen} shows the accuracy of the semantic detection by \sys{}. The results demonstrate that  \sys{} can still achieve  a comparable semantic  detection accuracy to the scenario-based experiments. This robustness is mainly due to the uniform nature of passengers' activities at railway station. Specifically,  most passengers follow similar routes, exhibit uniform behavior as they have a common target (boarding the train) starting from getting a ticket to boarding the train. In addition, \sys{} fuses multiple features (e.g., accelerometer, gyroscope and magnetometer based features are fused to identify the coin operated machines  as shown in the classification tree in Figure~\ref{other}) to identify the same semantic, reducing the sensitivity to specific  scenarios or machines.\\
\textbf{Per-Station Accuracy:}
To understand the classification accuracy on a per-station basis,
we show a one-station-out cross validation (one station data is selected as the validation data while all data collected at other stations is used as the database) results in Table \ref{cross}. Evident from the table, the classification accuracy does not deviate much between different stations due to fusing many features and the uniform railway passengers' behaviors. This further emphasizes the generalization ability of \sys{}.
\section{Discussion and Limitations}
\label{sec:dis}
We discuss some crowd-sensing challenges  addressed by the current version of \sys{}
along with our ongoing work.\\ \\
\noindent \textbf{System Robustness}\\
Our evaluation, quantified in Section~\ref{sec:eval}, is carried out by different users using different phones in different placement through different data collection methodologies and  spanned different stations and cities in Japan.  These extensive experiments verified \sys{} robustness under a wide range of scenarios. We believe that this robustness is based on a number of factors/design decisions including reorienting the phone sensors data, harnessing offset-invariant and orientation-invariant features, multi-modal sensor fusion, and combining a number of low level activities to identify the higher level passengers' activity. All allow \sys{} to generalize to different stations run by different operators, users, devices, and operation scenarios.\\\\
\noindent \textbf{Scalability}\\
We have tested \sys{} extensively in eight railway stations by 16 users under different scenarios. Scaling \sys{} to a worldwide scale is directly based on location clustering and leveraging the cloud. In particular, the semantics of each station can be processed independently of other stations based on the location of the collected traces. This spatial clustering lends itself nicely to the processing in the cloud, further enhancing the system scalability.\\\\
\noindent \textbf{Preserving User's Privacy}\\
 \sys{} gives users full control over their sensed data and  processes the  audio data  locally on the user's phone. Nonetheless,  some users may opt to  turn off  the sound or other sensors for privacy concerns. However, since \sys{} is a crowd-based system, it will still be able to identify these semantics from the samples submitted by other users.\\\\
\noindent \textbf{Handling Dynamic Changes}\\
The station internal  structures may evolve over time and, accordingly,  the state of the different  semantics (location changed or removed). To address this, \sys{} periodically rerun its clustering algorithm across time windows of different granularity to detect the removal of specific semantics. Specifically, the lack of a cluster with a specific size at the location of a previously detected semantic indicates the removal of this semantic. If clusters of correct samples of a semantic type formed in consecutive windows  are being mapped to a location that vary substantially from old instances locations, this is indicative of a change in the environment.  To classify the change type, we monitor newly uploaded  samples  of each instance of a semantic type. The change of location of a specific semantic is treated as a simple removal of this landmark and detecting it at a new location. We leave the evaluation of this aspect to future work due to space constraints.\\\\
\noindent \textbf{Other Indoor Environments}\\
Although \sys{} is designed for railway stations, it can  be customized to other indoor environments, e.g., airports, that have similar semantics (e.g., elevators,
stairs, escalators, vending machines, lockers, security gates, and automated boarding-pass printing machine).
Moreover, \sys{} can be extended to use semi-unsupervised learning techniques \cite{abdelnassersemanticslam,wang2012no} to extend to new environments without any prior
knowledge about what activities are expected. Specifically, new automatically learned semantic classes can be presented to a human user to provide the label to them, significantly reducing the overhead of extending \sys{} to new environments.
\section{Related Work}
\label{sec:related}
  The ideas in \sys{}  are built on three threads of research: mobile phone localization, human activity recognition and floorplan construction. We survey the most relevant work in each thread in the interest of space.\\

\noindent \textbf{Mobile Phone Localization} \\
Mobile phone localization has been well-studied with a variety of approaches so far \cite{youssef2015towards}. As GPS signal is not available  in many stations (e.g., subway stations),  an indoor localization  technique is needed to estimate the passenger's location. The most ubiquitous indoor localization techniques are either WiFi-based or dead-reckoning based. WiFi-based techniques, e.g., \cite{youssef2005horus,youssef2008horus,youssef2005multivariate,el2010propagation,youssef2006location,ibrahim2011hidden,eleryan2011synthetic}, require calibration to create a prior wireless map for the building. However, the calibration process is time consuming, tedious, and requires periodic updates; leading to the emergence of new calibration-free techniques \cite{secon}. Dead-reckoning based localization techniques, e.g., \cite{jin2011robust,wang2012no}, leverage the inertial sensors on mobile phones to dead-reckon the user starting from a reference point \cite{jin2011robust}. However, dead-reckoning error quickly accumulates leading to complete deviation from the actual path. Therefore,  many techniques have been proposed to reset the  dead-reckoning error including snapping to environment anchor points, such as elevators and stairs \cite{wang2012no,abdelnassersemanticslam}  and matching with the map information either indoor \cite{rai2012zee} or outdoor \cite{aly2013dejavu,aly2014map++}.\\
 \sys{} employs the basic   concept  in \cite{wang2012no,abdelnassersemanticslam} as it provides  accurate,  energy-efficient localization, and does not require an infrastructure support. However,  \sys{} discovers novel, activity-based, fine-grained and richer set of semantics targeting railway stations (e.g., vending machines, lockers, entrance gates, platforms, waiting (sitting) areas, among others) to reset the accumulated  localization error frequently.\\\\
\noindent \textbf{Human Activity Recognition} \\
Activity recognition literature has demonstrated the ability to recognize  user behavior using phone equipped sensors. Accelerometer data was used to detect the user transportation mode (walking, stationary, being in motorized transport){\cite{hemminki2013accelerometer,abdelaziz2015diversity,krumm2004locadio,reddy2010using,sohn2006mobility} 
 and it can classify more fine-grained activities and attributes like running, breathing rate, climbing up the stairs, biking, cleaning kitchen, vacuuming, and  brushing teeth  \cite{ravi2005activity,aly2016zephyr,abdelnasser2015ubibreathe,kwapisz2011activity,elhamshary2015activity}. Moreover,   accelerometer  data is  used  to detect more complex human activities like biking, lying, cleaning kitchen, cooking, sweeping, washing hands, and medication  \cite{dernbach2012simple}.  Ambient sensors like temperature, humidity, pressure, and light have been used to label user's location directly as being in kitchen, bedroom, bathroom and living room \cite{gerhard}. Moreover,  the AmbientSense system \cite{rossi2013ambientsense}  can recognize 23 different  contexts (e.g., coffee machine, raining, restaurant, dishwasher,  toilet flush, etc) by analyzing ambient sounds sampled from phone. In addition,  the RoomSense system in \cite{rossi2013roomsense}  uses  active sound probing to classify  the type of room (e.g., corridor, kitchen, lecture room, etc) where the user is  located. Ref. \cite{fan2014public}  actively probes the environment and  then analyzes the impulse response  on the phone to separate restroom from other rooms. Recently, RF-based device-free activity tracking and recognition has been used to detect different activities and the location of the person using standard RF networks \cite{youssef_keynote,kosba2009analysis,seifeldin2010deterministic}. \\
\sys{}  recognizes a higher level and novel set of  passenger activities  at railway stations. The crowd-sensed locations of
these activities are then mined to discover their uniquely associated stations semantics.\\\\
\noindent \textbf{Automatic Floorplan Construction} \\Recently,  a number of systems have been proposed that employ pedestrian motion traces to automatically construct indoor floorplans \cite{alzantot2012crowdinside, elhamshary2014checkinside,elhamshary2015semsense,gao2014jigsaw,jiang2013hallway,mine}. For instance, CrowdInside \cite{alzantot2012crowdinside} processes inertial motion traces using computational geometry techniques to extract the overall floorplan shape as well as corridors and room boundaries. It also identifies a variety of points of interest in the environment such as elevators and stairs.  However, their semantic detection method neither targets stations specific semantics (e.g., entrance gate, etc) nor it  provides  fine-grained classes of  elevation change semantics (e.g., stair types and elevator types). In addition,  their  elevator detection algorithm leverages only the motion pattern (Accelerate-Constant-Decelerate)   which may coincide with  normal  human walking patterns. This cannot happen in our method as normal walking traces are separated beforehand using the semantic type detection module. Finally, different  passengers' behaviors (e.g., climbing up or standing in  escalators) makes the low  acceleration variance, used in their method, is not a reliable discriminator between climbing  stairs and escalators. Jigsaw \cite{gao2014jigsaw} uses a computer vision approach to extract the position, size,  and orientation of landmark objects from images taken by users. It then combines user mobility traces and locations where images are taken  to produce the hallway connectivity and the room size. The system proposed in \cite{jiang2013hallway} leverages Wi-Fi fingerprints and user motion information to determine which rooms are adjacent in the building and estimating their sizes. It then orders them along each hallway and adjusts the room sizes to optimize the overall floorplan layout. \textit{Nevertheless, all previous systems did not attach any semantic information to the floorplan layout}. Finally, Ref. \cite{muralidharan2014barometric} assumes that there is a  difference in time to change floors using  elevators, escalators, and stairs and thus relies on the rate of height change (i.e., pressure)  to separate them. However,  this hypothesis  is neither  robust to different users walking speeds nor to different elevators/ escalators motion speeds.\\
The closest work to ours is  the  automatic  enrichment of indoor
floorplans  with semantic names  technique   in \cite {elhamshary2015semsense}.  
It  exploits phone sensors
data collected from users during their normal check-ins to
location-based social networks (LBSNs) (e.g., Foursquare) and combines them with
data extracted from the LBSNs databases to associate a place
name with its location on an unlabeled floorplan.  This system, however, can be applied only to indoor environments where the check-ins granularity level  matches the semantic names needed to enrich the map. For example, in shopping malls, user check-ins with venues names which can by leveraged by Semsense \cite {elhamshary2015semsense} to attach a  venue name to each room in the floorplan. However, users at railway stations usually do check-ins with stations names (they don't check-in using fine-grained station indoor semantics names)  which is not sufficient to infer  and locate the in-station semantics. \\
\sys{} assumes in its operation the availability of an unlabeled station floorplan by using one of these approaches. It then enriches the input floorplan with different semantics based on data collected from users' phones.
  \section{Conclusion}
  \label{sec:con}
We presented the \sys{} system for automatically enriching  transit indoor maps via a crowdsensing approach based on standard cell phones.  For energy efficiency, \sys{} leverages low-energy  phone sensors  and sensors that are already running for other purposes (e.g., inertial sensors).  We presented the \sys{}  architecture
as well as the features and classifiers that can accurately recognize different passenger's activities in railway stations which are  mined to  detect their uniquely  associated semantics.\\
 We implemented \sys{} using commodity mobile phones running the Android operating system and evaluated it at different railway stations in Japan. Our results show that \sys{} can detect  stations fine-grained semantics accurately  with  7.7\% false positive  and 7.5\% false negative rates on average leading to high accuracy in semantics location estimation. Finally, \sys{} can be  generalized over various stations running by different operators and user groups; and is robust to different phone placements while having a  significantly small energy
profile.\\
Currently we are expanding  \sys{} in multiple  directions  including inferring more station semantics,  handling dynamic changes in the environment, deployment of \sys{} in other indoor environments, among others.
\section{Acknowledgement}
We sincerely thank  the anonymous reviewers for their invaluable
feedback which helped improve the paper.\\
 This work is supported in part by JSPS KAKENHI grant numbers 26220001, 15H02690, and  26700006 to Osaka University and in part by a Google Research Award to E-JUST.
 
\bibliographystyle{abbrv}
\balance
\bibliography{mobisys_cam1}
%}
 \hfill{} \break

\end{document}